\documentclass[a4paper,10pt,english]{article}
\usepackage[margin=1in]{geometry}
\usepackage[utf8]{inputenc}
\usepackage{enumitem}
\usepackage[colorlinks]{hyperref}
\usepackage{graphicx}
\usepackage{color}
\usepackage{amsmath}
\usepackage{amssymb}
\usepackage{enumitem}
\usepackage{authblk}
\usepackage{xcolor}
\usepackage{gensymb}
\usepackage{setspace}
\usepackage[
backend=biber,
style=numeric,
sorting=none
]{biblatex}
\title{Quantum Technology for Military Applications}
\author[1,2]{Michal Krelina\thanks{michal.krelina@cvut.cz}}
\affil[1]{Faculty of Nuclear Sciences and Physical Engineering, Czech Technical University in Prague, Brehova 7, Prague, Czech Republic}
\affil[2]{Quantum Phi s.r.o., Bryksova 944, Prague, Czech Republic}

\date{}

\begin{document}
\maketitle

\begin{abstract}
Quantum technology is an emergent and potentially disruptive discipline, with the ability to affect many human activities.
Quantum technologies are dual-use technologies, and as such are of interest to the defence and security industry and military and governmental actors.
This report reviews and maps the possible quantum technology military applications, serving as an entry point for international peace and security assessment, ethics research, military and governmental policy, strategy and decision making.  
Quantum technologies for military applications introduce new capabilities, improving effectiveness and  increasing precision, thus leading to `quantum warfare', wherein new military strategies, doctrines, policies and ethics should be established.
This report provides a basic overview of quantum technologies under development, also estimating the expected time scale of delivery or the utilisation impact.
Particular military applications of quantum technology are described for various warfare domains (e.g. land, air, space, electronic, cyber and underwater warfare and ISTAR---intelligence, surveillance, target acquisition and reconnaissance), and related issues and challenges are articulated.
\end{abstract}

%\begin{keyword}
\textbf{Keywords:} quantum warfare, quantum technology, quantum computing, quantum sensing, quantum network, quantum radar, quantum imaging, military applications, quantum security, dual-use technology
%\end{keyword}

%---------------------------------------------------------------------------
%---------------------------------------------------------------------------

\section{Introduction}

Although fourth generation modern warfare is characterised by decentralisation and the loss of states' monopoly on war \cite{william1989changing,Lind2004UnderstandingFG}, armies of advanced countries characteristically have access to state-of-the-art military technologies. This includes the appearance of quantum technologies on the horizon.

The term quantum technology (QT) means the technology mostly arising out of the so-called second quantum revolution. Earlier, the first quantum revolution brought technologies that are familiar to us today, such as nuclear power, semiconductors, lasers, magnetic resonance imaging, modern communication technologies or digital cameras and other imaging devices. 
The first quantum technology resulted in nuclear weapons and energy; then, the classical computer gained a significant role. Presently, laser weapons are being implemented and tested \cite{AFFANAHMED2020}.

The second quantum revolution \cite{Dowling2003} is characterised by manipulating and controlling individual quantum systems (such as atoms, ions, electrons, photons, molecules or various quasiparticles), allowing to reach the standard quantum limit; that is, the limit to measurement accuracy at quantum scales. In this report, the term quantum technology refers to the technology from the second quantum revolution. Quantum technology does not bring fundamentally new weapons or standalone military systems, but rather significantly enhances measurement capability, sensing, precision and computation power and efficiency of the current and future military technology. Most of the quantum technologies typically are technologies of dual use. Consequently, there is tremendous potential for military applications of quantum technology.
Various studies and recommendations are emerging, signalling the increasing likelihood of such technology being realised; see, for example, \cite{UKLandscape2016,ASPI2017,IDA2019,FFI2020}.

This report provides a more in-depth context in which to understand the term `quantum warfare', discussing the possibility of its affecting the intelligence, security and defence sectors, and describing new possible capabilities or improvements. The goal is not to provide a precise forecast of quantum-based technologies, but rather to show possible directions and trends in implementation and applications. Quantum technologies in general are considered emerging technologies, with the potential to change the conduct of warfare and the outcomes of battles \cite{FFI2020}. Although the current quantum technologies mostly have low Technology Readiness Levels (TRL), they are believed to have disruptive potential \cite{NQIT2018}.
The mapping of quantum technologies' conceivable military applications is also important for the further assessment of threats to global peace and in the discussion of ethics policies or quantum-based preventive arms control.

This report comprises eight sections.
In Section~\ref{sec:definitions}, the terms `quantum technology' and `quantum warfare' are defined, with the quantum technology taxonomy and quantum technologies being introduced. 
Section~\ref{sec:overview} provides the basic quantum technology overview that is the foundation for a particular application, including the expected time of deployment and utilisation impact. 
Section~\ref{sec:strategy} presents the general considerations and expectations regarding quantum technology development and deployment in the military domain.
In Section~\ref{sec:warfare}, the applications of individual quantum technologies in the military are presented for various domains ( e.g. cyber, underwater, space and electronic warfare).
Section~\ref{sec:optipess} identifies and discusses the quantum hype as well as the realistic possibilities. 
Section~\ref{sec:consandchall} contains the initial discussion on related military, peace and ethical aspects as well as the technical consequences and challenges.
Section~\ref{sec:conclusions} concludes the paper.

Sections~\ref{sec:warfare} and \ref{sec:strategy} concern national security and defence issues. 
Although Section~\ref{sec:overview} is based on state-of-the-art research and provides related references, Section~\ref{sec:warfare} is based more on various military or government reports, policy briefs and international security analyses such as \cite{UKLandscape2016,ASPI2017,IDA2019,FFI2020,ATARC,AusQRoadmap,DefensePrimer,DSTL-QC}.
Here, the reader should be wary of the hype surrounding quantum technology  and avoid exaggerated expectations; this aspect is addressed in Section~\ref{sec:optipess} and by \cite{Biercuk2020war}. 
For many of the presented quantum technology military applications, it is uncertain whether all challenges connected with the demands of high-end military technologies will be resolved, or even that the technology will actually be deployed.

%---------------------------------------------------------------------------
%---------------------------------------------------------------------------
%---------------------------------------------------------------------------
\section{Definitions}\label{sec:definitions}

The term quantum technology is defined as follows:
\begin{quote}
\textit{Quantum technology} (QT) is an emerging field of physics and engineering based on quantum-mechanical properties---especially quantum entanglement, quantum superposition and quantum tunnelling---applied to individual quantum systems, and their utilisation for practical applications.    
\end{quote}
As follows from the definition, quantum technology describes the various physical principles of quantum-mechanical systems, with numerous applications; for instance, the technique of trapped ions can serve as a quantum bit for quantum computers and as a quantum sensor for magnetic fields or quantum clocks.
\begin{quote}
\textit{Dual-use technology} refers to fields of research and development with potential application in both defence and commercial production \cite{NATO-dualuse}.     
\end{quote}
Quantum technology is a typical dual-use technology which has been of considerable interest not only for military but also for governmental actors \cite{nouwens_legarda_2018} and peacekeeping organisations. 
%The dual-use nature means that industry and academia will also play key roles in inventing and adapting these new technologies \cite{military_2019}. 

\begin{quote}
\textit{Quantum warfare} (QW) is warfare that uses quantum technologies for military applications that affect intelligence, security and defence capabilities of all warfare domains, and it ushers in new military strategies, doctrines, scenarios and peace as well as ethics issues.    
\end{quote}
There have been attempts also to define the \textit{quantum domain} \cite{Davidson2019} as a new domain for warfare. However, in this text, we will consider quantum technology as a factor that improves all currently defined domains, rather than as a standalone warfare domain. 

Subsequently, it is helpful to define the term \textit{quantum attack}, which refers to using quantum technologies to break, disrupt or eavesdrop on either classical or quantum security systems. Typical examples are eavesdropping using quantum key distribution or quantum computers breaking the Rivest–Shamir–Adleman (RSA) encryption scheme. 

Although there is plenty of QT literature, there is no explicit agreement on quantum technology taxonomy. 
We will use the following taxonomy:
\begin{itemize}[noitemsep]
    \item \textbf{Quantum Computing and Simulations}
    \begin{itemize}
        \item Quantum Computers (digital and analogue quantum computers and their applications, such as quantum system simulation, quantum optimisation, ...)
        \item Quantum Simulators (non-programmable quantum circuits)
    \end{itemize}
    \item \textbf{Quantum Communication and Cryptography}
    \begin{itemize}
        \item Quantum Network and Communication (quantum network elements, quantum key distribution, quantum communication)
        \item Post-Quantum Cryptography (quantum-resilient algorithms, quantum random number generator)
    \end{itemize}
    \item \textbf{Quantum Sensing and Metrology}
    \begin{itemize}
        \item Quantum Sensing (quantum magnetometers, gravimeters, ...)
        \item Quantum Timing (precise time measurement and distribution)
        \item Quantum Imaging (quantum radar, low-SNR imaging, ...)
    \end{itemize}
\end{itemize}

Aside from the general quantum technology taxonomy presented above, we introduce a new division of quantum technologies according to their benefits and utilisation. The following classification can be generalised; however, we place more emphasis on military applications.
The quantum technology utilisation impact classification is as follows:
\begin{itemize}
    \item \textbf{Must have}: quantum technology that has to be implemented to protect against future quantum attacks (e.g. post-quantum cryptography);
    \item \textbf{Effectiveness}: quantum technologies that increase the effectiveness of the current technology and methods (e.g. quantum optimisations, quantum machine learning or artificial intelligence);
    \item \textbf{Precision}: quantum technologies that increase the precision of the current measurement technology (e.g. quantum magnetometry, quantum gravimetry, quantum inertial navigation, timing);
    \item \textbf{New capabilities}: quantum technologies offering new capabilities that were beyond the scope of the present technology (e.g. quantum radar, quantum simulation for chemistry, quantum cryptoanalysis, quantum key distribution).
\end{itemize}
Note that this classification is not mutually exclusive.

%---------------------------------------------------------------------------
%---------------------------------------------------------------------------
%---------------------------------------------------------------------------
\section{Quantum Technology Overview}\label{sec:overview}

This section provides a basic description of quantum technologies, with related references.  
For each quantum technology, the current development status is presented, the utilisation impact determined, expected time to deployment estimated\footnote{
Short-term: 0--5 years,
Mid-term: 6--10 years, 
Long term: 10--20 years.
}, and the main challenges are sketched. For quantum computing application, the approximate number of required logical qubits is provided.

Different quantum technologies and their applications are at different TRLs\footnote{See \href{https://ec.europa.eu/research/participants/data/ref/h2020/wp/2014_2015/annexes/h2020-wp1415-annex-g-trl_en.pdf}{Technology Readiness Levels according to EU}, \url{https://ec.europa.eu/research/participants/data/ref/h2020/wp/2014_2015/annexes/h2020-wp1415-annex-g-trl_en.pdf}} from TRL 1 (e.g., some types of qubits) to TRL 8 (e.g., quantum key distribution).

We are not aiming for completeness here, nor do we present any theoretical background, but just introduce the basics, the effects and the current state of development, as needed to follow the discussed military applications.

%---------------------------------------------------------------------------
%---------------------------------------------------------------------------
\subsection{Quantum Information Science}
Quantum information science (QIS) is an information science related to quantum physics, and deals with quantum information. In classical information science, the elementary carrier of information is a bit that can be only 0 or 1. The quantum information elementary carrier of information is the quantum bit, qubit in short. A qubit can be $|0\rangle$  or $|1\rangle$, or an arbitrary complex linear combination of states $|0\rangle$ and $|1\rangle$ called the quantum superposition. 

The other crucial property is the quantum entanglement. Quantum entanglement refers to a strong correlation between two or more qubits (or two or more quantum systems in general) with no classical analogue. Quantum entanglement is responsible for many quantum surprises. Another feature is the no-cloning theorem \cite{Park1970}, which says that quantum information (qubit) cannot be copied. This theorem has profound consequences for qubit error correction as well as for quantum communication security. 

Quantum information science describes the quantum information flow in quantum computing and quantum communications, although in a broader sense it can be applied in quantum sensing and metrology, see \cite{nielsen2010quantum,mermin2007quantum}. 

There is considerable academic interest, and several quantum algorithms have been created \cite{QA-ZOO}. However, only a few are expected to be valuable for defence and security applications.

%---------------------------------------------------------------------------
%---------------------------------------------------------------------------
\subsection{Quantum Computing}
\begin{itemize}[noitemsep]
\item \textbf{Status: } commercially available with very limited number of physical qubits
\item \textbf{Utilisation impact: } new capability, effectiveness, precision
\item \textbf{Timeline expectation: } one million physical qubits in ten years
\item \textbf{Main challenges: } improving the quality of qubits (coherence, error resistance, gate fidelity), upscaling the number of qubits, logical qubits 
\end{itemize}

Quantum computing refers to the utilisation of quantum information science to perform computations.
Such a machine can be called a quantum computer. 
The classification of quantum computers can be very complex. For the purposes of this report, we simplify the classification as follows:
\begin{itemize}
    \item \textit{Digital quantum computer} (also called a gate-level quantum computer) is universal, programmable and should perform all possible quantum algorithms and have numerous applications described below. Classical computers can fully simulate the gate-level based quantum computer. The difference is in resources and speed. For instance, the simulation of fully entangled qubits increases the requirement of classical resources exponentially. This means that the simulation of $\gtrsim 45$ qubits is practically impossible on the classical (super)computers.
    
    \item \textit{Analogue quantum computer} (also called Hamiltonian computation) is usually realised using quantum annealing (as the noise version of the adiabatic quantum computing). Quantum annealer differs from the digital quantum computer by the limited connectivity of qubits and different principles.
    Therefore, the utilisation of analogue quantum computers is more constrained but is still suitable for tasks such as quantum optimisations or Hamiltonian-based simulations.

    \item \textit{Quantum simulator} is used for the study and simulation of other quantum systems that generally are less accessible and is usually built as a single-purpose machine. In comparison with a quantum computer, the quantum simulator can be imagined as a non-programmable quantum circuit. 
\end{itemize}

In general, quantum computing will not replace classical computation.
Quantum computers will be practical and useful for a limited type of problems only, typically problems with high complexity. 
The actual deployment of quantum computing applications depends on the quality (coherence, error resistance, gate fidelity) and the number of qubits.
Some of the essential parameters to follow are: the number of qubits, qubit coherence time, quantum-gate fidelity and qubit inter-connectivity. 
The set of quantum instructions applying quantum gates on individual qubits is called a quantum circuit. A quantum circuit is a practical realisation of the quantum algorithm.

Following \cite{IDA2019}, quantum computers can be classified into three evolutionary stages: Component quantum computation (CQC), Noisy intermediate-scale quantum (NISQ) computing and Fault-tolerant quantum computing (FTQC).
The CQC stage covers quantum computing demonstrators and maturing the basic elements. CQC has a very limited computational capability that is sufficient for demonstration of some proofs of principle.
The NISQ stage quantum computer should have a sufficient number of qubits to demonstrate the advantages of quantum computing. Continuous research should lead to increasing the number and quality of qubits.
The FTQC stage starts when a perfect logical qubit is reached (for an explanation, see below).

Physical qubits can be realised by numerous quantum systems. The most recent advanced are quantum computer based on superconducting qubits and the trapped-ion qubits that are in or close to the NISQ stage. All other technologies, such as cold atoms, topological, electron spin, photonic or NV centre-based qubits, are still in CQC stage or theory only.
The individual quantum computers and their performance differ significantly (in, e.g. speed, coherence time, the possibility to entangle all qubits, gate fidelity). Various metrics and benchmarks, such as Quantum Volume metric \cite{Cross2019}, have been developed for their comparison.

The problem, common to all types of qubits, is their quality. A qubit is very fragile and has a limited coherence time (a time scale during which it will not lose the quantum information). Every operation performed on a qubit has limited fidelity. Researchers accordingly need to employ the error correction codes. The error correction for qubits is much more complicated than error correction of classical bits, because qubits cannot be copied, as the no-cloning theorem explains. 
Two types of qubits are distinguished: the physical qubit realised by a physical quantum system and the logical qubit consisting of several physical qubits and error correction codes. A logical qubit is a perfect or near-perfect qubit with very long-to-infinity coherence time, very high fidelity and higher environment resistivity. 
For example, based on surface error correction protocol, for one logical qubit, depending on the algorithm, up to 10 000 physical qubits \cite{Fowler2012} will be needed. 
For a recent overview of quantum computing, see, for example, \cite{national2019quantum}.

Examples of leading-edge quantum computers are the quantum computer with 53 physical superconducting qubits, manufactured by Google (which claimed quantum supremacy in 2019 \cite{Arute2019}), and the one by IBM.
The best trapped-ion quantum computers are of 32 qubits by IonQ or six qubits by Honeywell. In the case of photonic qubits, there is a 24-qubit quantum computer by Xanadu.
The anticipated timelines as imagined by the quantum computing roadmap by IBM and Google are the following:
IBM plans a 433-qubit quantum processor in 2022 and 1,121 qubits by 2023 \cite{IBMroadmap}. Google has announced a plan to reach a quantum module of 10,000 qubits. All other quantum processors would consist of such modules up to 1 million qubits in 2029 \cite{Google1million}.
Based on a survey among leaders in key relevant areas of quantum science and technology, it is likely that quantum computers will start to become powerful enough to pose a threat to most of the public key encryption schemes (for more details, see Sec.~\ref{sec:cryptoanal}) in about two decades \cite{Mosca2019}.
Examples of analogue quantum computers are the quantum annealer by D-Wave Systems with over 5,000 qubits and the coherent Ising machine by Toshiba.

The difference between analogue and digital quantum computers lies in their different physical principles and their limitations.
The digital quantum computer is limited by resources and not by noise (noise can be corrected using more resources). In contrast, the analogue quantum computer is limited by noise which is difficult to understand, control and characterise (especially for a quantum annealer). Therefore, analogue quantum computers' applicability is limited \cite{national2019quantum}.

In reality, the tasks accomplished by quantum computers will be mostly only subprograms or subroutines of the classical computer programs. The classical program will not only control quantum computers but will also provide a lot of computation that it would be impractical to carry out on a quantum computer. This includes the recent applications of quantum simulation in chemistry using, for example, the Variational quantum eigensolver (VQE) \cite{QiskitVQE}, which is a hybrid combination of classical and quantum computing. Also, quantum computers are large machines, many of which require cryogenics. It is unlikely therefore that in the decades to come most customers will acquire a personal quantum computer, but rather they will access these as a service in the cloud. 
The cloud-based models of quantum computing (often called Quantum Computing-as-a-Service – QCaaS) are commercially available nowadays, even for free, and they allow access to anyone interested in quantum computing.
The cloud access to quantum computers is offered by individual quantum hardware manufacturers. Some platforms, such as Microsoft Azure Quantum or Amazon Braket, offer access to quantum computers of various manufacturers within one ecosystem.

It is also helpful to clarify the terms of quantum supremacy, advantage and practicality. 
\textit{Quantum supremacy} is a case where a particular problem is solved by a quantum computer significantly faster than by a classical computer. However, the problem is likely to be theoretical rather than practical. 
\textit{Quantum advantage} refers to a case when a quantum computer is able to solve real-world problems that classical computers cannot.
\textit{Quantum practicality} is similar to quantum advantage, with the only difference being that the quantum computer solves real-world problems faster than the classical computer.

We provide below a basic overview of possible quantum computer applications. The reader should keep in mind that quantum computing is a fast developing sector, and new revolutionary quantum algorithms are still waiting to be discovered.
Note that, in the context of quantum computing applications, the term `qubit' implies a logical qubit. However, small quantum circuits can be run with only physical qubits, with reasonable precision.

%...................................................................................
\subsubsection{Quantum Simulations} \label{sec:quantumSimulations}
\begin{itemize}[noitemsep]
\item \textbf{Status: } algorithms in development, small-scale applications
\item \textbf{Utilisation impact: } new capability (e.g. quantum chemistry computation)
\item \textbf{Timeline expectation:} short-term, usability scales up with the number of qubits
\item \textbf{Qubits requirement: } $\sim 200$ (e.g. for nitrogen fixation problem)
\item \textbf{Main challenges: } number of logical qubits
\end{itemize}

Long before the first quantum computer was created, the main task for the quantum computer was considered the simulation of other quantum systems \cite{feynman_simulating_1982}. Molecules are such a quantum system. Despite advancement of extant computing power, the full simulation of only simpler molecules can be performed using present computational chemistry, or of larger molecules for the price of many approximations and simplifications. 
For example, for a system with $n$ electrons, the classical computer would need $2^n$ bits to describe the state of electrons, whereas the quantum computer needs only $n$ qubits. 
Therefore, quantum simulations are the first and still perhaps the most promising application for quantum computers.

The most dominant approaches are two: quantum-phase estimation \cite{Reiher2017} and quantum variational techniques (VQE) \cite{Peruzzo2014,McClean2016}. The latter approach in particular has the highest likelihood of success on NISQ computers; for example, in 2020, Google performed the biggest quantum chemical simulation up to date (of H$_{12}$ molecule using the VQE) \cite{Arute2020}.

Algorithms for quantum chemistry simulations are being developed. They can be applied to more complex simulations, hand in hand with the number of qubits. Therefore, even at this early stage of quantum computing, there is significant interest from the chemical and pharmaceutical industries.
In general, such simulations allow the discovery and design of new drugs, chemicals and materials.
Recently considered topics, for instance, are high-temperature superconducting, better batteries, protein folding, nitrogen fixation and peptides research.

%...................................................................................
\subsubsection{Quantum Cryptoanalysis}\label{sec:cryptoanal}
\begin{itemize}[noitemsep]
\item \textbf{Status: } algorithms ready
\item \textbf{Utilisation impact: } new capability (e.g. public-key cryptography schemes breaking)
\item \textbf{Timeline expectation:} mid- to long-term
\item \textbf{Qubits requirement: } $\sim 6,200$  for 2048 bit RSA factorisation \cite{Gidney_2021}, $\sim 2,900$ for 256 bit ECDLP-based\footnote{
Cryptography based on Elliptic Curve Discrete Logarithm Problem.
} encryption \cite{Hner2020}
\item \textbf{Main challenges: } number of logical qubits
\end{itemize}

One of the most well-known quantum computer applications is the factorisation of large prime numbers by exponential speedup described by Shor's algorithm \cite{Shor1994}. This is a threat for public-key cryptography schemes, such as RSA, DH and ECC\footnote{cryptography schemes named after Rivest–Shamir–Adleman, Diffie–Hellman, Elliptic Curve Cryptography}, based on the large prime number multiplications, the discrete logarithm problem or the elliptic-curve discrete logarithm problem-based schema that are considered computationally intractable or very hard for classical computers. 

Although the resources of existing NISQ quantum computers are far from what is needed for RSA breaking, the threat is quite real. An adversary or foreign intelligence could intercept and store encrypted traffic until the quantum cryptoanalysis becomes available. Because the time of declassification of many secrets is far beyond the expected timelines for powerful quantum computer delivery, such a threat can be considered genuine, nowadays.

Quantum cryptoanalysis also offers improved tools for a brute-force attack on the symmetric encryption schemes. For example, the well-known Grover's searching algorithm \cite{Grover1996} reduces the key security by half against a brute-force attack; a 256-bit AES\footnote{Advanced Encryption Standard} key could be resolved by brute force in roughly $2^{128}$ quantum operations. Despite the huge resource requirements of quantum computers, doubling the symmetric key length \cite{Bernstein2010}is recommended, nevertheless.
Moreover, Simon's algorithm and superposition queries \cite{Simon2002} can completely break most message authentication code (MAC), and authenticated encryption with associated data (AEAD), such as HMAC-CBC and AES-GCM\footnote{Hash-based Message Authentication Code-Cipher Block Bhaining, AES with Galois/Counter Mode} \cite{kaplan2016breaking,bonnetain2020quantum}.

Further, there is active research on cryptoanalytic attacks upon symmetric key systems based on the structure present in symmetric cryptosystems, which can offer up to super-polynomial speedup \cite{Kaplan2016}. However, these algorithms suffer from excessive resource requirements on the quantum computer.

%...................................................................................
\subsubsection{Quantum Searching and Quantum Walks}
\begin{itemize}[noitemsep]
\item \textbf{Status: } algorithms under development
\item \textbf{Utilisation impact: } effectiveness (e.g. faster searching)
\item \textbf{Timeline expectation:} short- to mid-term
\item \textbf{Qubits requirement: } $\sim 100$, depends on the searched system size
\item \textbf{Main challenges: } number of logical qubits
\end{itemize}

One of the most famous searching quantum algorithms is the Grover's algorithm \cite{Grover1996}, which offers quadratic speedup in database searching, or generally in inverting a function. For an unsorted list or database, the classical searching algorithms are about complexity $\mathcal{O}(N)$ (meaning proportional to the number of $N$ entities), although Grover's algorithm is about $\mathcal{O}(\sqrt{N})$.

Quantum searching algorithms are an important topic for the so-called Big Data (unstructured data) analysis. Working on a large amount of data requires a large quantum memory. However, there is no reliable quantum memory that would keep the quantum information for an arbitrarily long time and in large amounts. Second, the transformation of classical data to the quantum form is time-consuming and ineffective. Therefore, only the searching on data generated algorithmically is considered practical at the moment.

The other approach to searching can be based on the quantum random walk mechanism \cite{Shenvi2003}, which offers similar speedup as Grover's algorithm.

%...................................................................................
\subsubsection{Quantum Optimisations}
\begin{itemize}[noitemsep]
\item \textbf{Status: } algorithms in development
\item \textbf{Utilisation impact: } effectiveness (e.g. faster solution of NP problems)
\item \textbf{Timeline expectation:} short- to mid-term
\item \textbf{Qubits requirement: } $\sim$ 100, depends on the problem complexity 
\item \textbf{Main challenges: } number of logical qubits
\end{itemize}

Quantum optimisation is a very actively explored topic, given the possibility of solving NP-level\footnote{NP is a complexity class characterised by the fact that it cannot be solved in polynomial time but can be verified in polynomial time. Specifically, the NP-hard problems are not only hard to solve but are difficult to verify as well. Examples of NP-hard problems are the Travelling Salesman Problem and Graph Colouring problems.} complex problems. 
An example of such an NP problem is the travelling salesman problem.
Here, given a list of places and the distances between them, the goal is to find the shortest (and optimal) route. Naively, one can try all possibilities, but such an approach has severe disadvantages, and may even become impossible, with increasing complexity.
Therefore, the most common solutions are based on heuristic algorithms that are not necessarily guaranteed to find the best solution but at least one close to it.

Quantum computing introduces a new perspective on the issue and offers different approaches and techniques. The most dominant methods are currently based on a variational approach, such as the Quantum approximate optimisation algorithm (QAOA) \cite{farhi2014quantum}. Part of QAOA is the sub-technique called Quadratic unconstrained binary optimisation (QUBO) \cite{glover2019tutorial}, which is also suitable for analogue quantum computers. Other methods are, for example, the quantum analogy of the least-squares fit \cite{Wiebe2012}  or semidefinite programming \cite{brandao2017quantum}.

So far, it is not clear whether the quantum optimisation will offer some speedup against the classical heuristic methods. However, there is consensus that if at all some speedup is achievable, it will not be more than polynomial \cite{brandao2017quantum}.
A new paradigm introduced by quantum computing leads to new quantum-inspired classical algorithms, such as in the case of QAOA \cite{barak2015beating} that delete the quantum speedup. On the other hand, we can speak about quantum-inspired algorithms as of the first quantum computing practical result.

There have been many demonstrations, use cases and proofs of concept for quantum optimisations, especially in connection with analogue quantum computing that currently offers the most quantum computing resources for such applications. The typical demonstrations were optimisations for traffic, logistics or the financial sector\footnote{For examples of developed quantum optimisation applications at D-Wave's quantum annealer, see \url{https://www.dwavesys.com/applications}.}.

%...................................................................................
\subsubsection{Quantum Linear Algebra}\label{sec:linalg}
\begin{itemize}[noitemsep]
\item \textbf{Status: } algorithms in development
\item \textbf{Utilisation: } effectiveness (e.g. faster linear equation solving)
\item \textbf{Timeline expectation:} short- to mid-term
\item \textbf{Qubits requirement: } depends on the solved system size
\item \textbf{Main challenges: } number of logical qubits
\end{itemize}

It has been shown that quantum computers can reach super-polynomial speedup for solving linear equations also. Such a speedup was estimated especially for the HHL (Harrow-Hassidim-Lloyd) \cite{Harrow2009} algorithm for sparse matrices. However, the estimated speedup depends on the size of the problem (matrix). There are also large resource requirements, which for some problems can be considered too impractical \cite{Scherer2017}. On the other hand, for the system of linear equations of 10,000 parameters, for instance, 10,000 steps are needed to solve it, whereas the HHL can provide an approximate solution after 13 steps.

At present, many numerical simulations in planning, engineering, construction and weather forecasting simplify complex problems as a large set of linear equations. For many of them, being statistical in nature, the approximated solution could be sufficient. 

Note that the HHL algorithm was shown as universal for quantum computing and was demonstrated for various applications such as $k$-mean clustering, support vector machines, data fitting, etc. For more details, see \cite{Aaronson2015}.

One of the major caveats of quantum algorithms working with a large amount of input data is data loading. Classical data, especially binary data or bits, need to be transferred into quantum states for follow-on processing by efficient quantum algorithms. This process is slow, and the classical data loading itself can take longer than the coherence time. The solution is a quantum memory or quantum RAM \cite{Aaronson2015,Blencowe2010}.

%...................................................................................
\subsubsection{Quantum Machine Learning and AI}
\begin{itemize}[noitemsep]
\item \textbf{Status: } algorithms in development
\item \textbf{Utilisation impact: } effectiveness (e.g. better machine learning optimisations )
\item \textbf{Timeline expectation:} short- to mid-term
\item \textbf{Qubits requirement: } $\sim 100$, depends on the problem complexity 
\item \textbf{Main challenges: } number of logical qubits
\end{itemize}

Due to the hype around classical machine learning and artificial intelligence (ML/AI), it can be expected that there will be quantum research on this topic also.
First, note that one cannot expect full quantum ML/AI, considering the very low efficiency of working with classical data \cite{Biamonte2017}, all the more so if the missing quantum memory and very slow loading and coding of classical data (e.g. data of picture) into quantum information format are considered. It is simply not practical. Another situation will emerge when ML/AI is applied to quantum data; for instance, from quantum sensors or imaging \cite{Dunjko2016}.

Nevertheless, quantum-enhanced ML/AI \cite{wittek2016quantum,Dunjko2020nonreviewofquantum} can be introduced, where quantum computing can improve some machine learning tasks such as quantum sampling, linear algebra (where machine learning is about the processing of complex vectors in a high-dimensional linear space) or quantum neural networks \cite{Biamonte2017}. 
One example is the quantum support vector machine \cite{Havlek2019}.

In fact, the ML/AI topic covers various techniques and approaches, and it is no different in connection with quantum computing. Quantum ML/AI or quantum-enhanced ML/AI is the subject of many research works nowadays. For a survey of quantum ML/AI algorithms and their possible speedup, see \cite{Ramezani2020}.

%---------------------------------------------------------------------------
%---------------------------------------------------------------------------
\subsection{Quantum Communication and Cryptography}

Quantum communication refers to a quantum information exchange via a quantum network that uses optical fibre or free-space channels. In most cases, quantum communication is realised using a photon as the quantum information carrier. However, due to the limitations of photons, such as losses at large distances, the quantum network contains other elements such as a quantum repeater or quantum switch.

The goal of quantum cryptography is to replace conventional (mainly asymmetric) encryption schemes with quantum-resistant algorithms with the quantum key distribution. The typical quantum features used for quantum communication are the following: quantum entanglement, quantum uncertainty, and the no-cloning theory which states that quantum information cannot be copied \cite{Park1970,Gisin2007}.

%...................................................................................
\subsubsection{Quantum Network}\label{sec:qunet}
\begin{itemize}[noitemsep]
\item \textbf{Status: } in research (commercially available for QKD with trusted nodes only) 
\item \textbf{Utilisation impact: } new capability, effectiveness (e.g. ultra secure communication, quantum-resilient cryptography)
\item \textbf{Timeline expectation:} mid-term
\item \textbf{Main challenges: } quantum repeater and switch (quantum memory)
\end{itemize}

The aim of the quantum network (sometimes called quantum internet \cite{Wehner2018} or quantum information network (QIN)) is to transmit quantum information via numerous technologies across various channels.
The quantum information (qubit) is usually carried by individual photons, and as such the quantum information transmission is fragile. Moreover, many quantum network applications rely on quantum entanglement. 

The usual channels for quantum information transmission are specialised low-loss optic fibres or the current telecommunication optic fibre infrastructures with higher loss.
The case of two communicating endpoints close to each other is as simple as using one optical fibre. The complexity of the network increases with more end nodes or large distances, where components such as a quantum repeater or quantum switch are required. Note that very modest (one qubit) quantum processors are sufficient for most quantum network applications.

The free-space quantum channel is more challenging. Optical or near-optical photons are of limited utility in the atmosphere due to the strong atmospheric attenuation. 
Therefore, the most commonly considered and realised quantum network scenario is using quantum satellites \cite{Yin2017,Yin2017v2}.
The advantage of satellites is the possibility of utilising optical-photon communication for transmission of the quantum information, where the losses in the satellite--ground link are lower than the loss between two ground nodes far apart.
Nevertheless, the optical photons' communication in the free-space channel for short distances can be realised using, for instance, drones \cite{Liu2021}. The best way would be to use the microwave spectrum as employed by classical wireless communication. However, communication that uses the microwave spectrum at the level of individual photons is even more challenging \cite{Pogorzalek2019}. Microwave single-photon technology involves greater difficulty in generating and detecting individual photons. Another problem is a noisy environment in microwave bands.

Quantum communication at long distances requires quantum repeaters due to photon loss and decoherence. A quantum repeater is an intermediate node that works similarly to the amplifier in classical optical networks but needs to obey the no-cloning theorem. In fact, the quantum repeater allows entangling qubits of end nodes. 
When two end nodes are entangled, the effect of quantum teleportation \cite{Bouwmeester1997} can be exploited.
This means that the quantum information can be teleported without a physically sent photon; just a classical communication is required. 
Utilising quantum entanglement, the quantum information can then flow through a quantum network or part of it, which can even be under eavesdropper control without any chance of revealing the transmitted quantum information. For correct functioning of the quantum repeater, quantum memory is required. However, no reliable and practical quantum memory is available yet.

As an intermediate step, a trusted repeater can be used. The trusted repeater will not entangle end nodes and is used for the quantum key distribution (QKD, see the next section, \ref{sec:QKD}) only. To imagine how it works, let us consider two parties $A$ and $B$ and a trusted repeater $R$. Then the key $k_{AB}$ is encrypted with key $k_{AR}$. The trusted repeater $R$ decrypts $k_{AR}$ to get $k_{AB}$. At this point, the trusted repeater $R$ knows the key $k_{AB}$, and $A$ and $B$ have to trust that the key is safe and not under the control of the eavesdropper. Finally, $R$ re-encrypts $k_{AB}$ using the $k_{RB}$ key and sends it to $B$. This is a technique used in present QKD networks.

The next step, currently tested in experiments, is the measurement device-independent QKD (MDI-QKD) \cite{Braunstein2012,Lo2012}. It is a quantum protocol that not only replaces trusted repeaters (still not quantum, no support of entanglement) with secure repeaters, but also serves as a switch. That means the usual star network topology and infrastructure can start to be built. Note that in the MDI-QKD network, attacks on the central node physically cannot reveal the key nor reveal sensitive information. Later, the central nodes will be replaced by the quantum switch and repeater, and the fully functional quantum information network will be achieved.

Quantum networks will work in parallel with the classical networks, since not all transmitted information needs to be encoded in quantum information.
In fact, parallel classical network is required, for instance, for quantum teleportation.
Quantum networks can be used for the following applications:
\begin{itemize}
\item[-] \textit{Quantum key distribution (QKD)}, a secure transmission of cryptographic key (see Section~\ref{sec:QKD});
\item[-] \textit{Quantum information transmission} between quantum computers or quantum computing clusters at large distances or for sharing of remote quantum capabilities;
\item[-] \textit{Blind quantum computing} \cite{Broadbent2009,Fitzsimons2017} allowing to transmit a quantum algorithm to quantum computer, perform computations and retrieve results without the owner or eavesdropper knowing what the algorithm or result was;
\item[-] \textit{Network clock synchronisation} \cite{Chuang2000}, see Section~\ref{sec:clocks};
\item[-] \textit{Secure identification} \cite{Damgrd2014} allowing identification without revealing authentication credentials;
\item[-] \textit{Quantum position verification} \cite{Unruh2014} allowing to verify the position of the other party;
\item[-] \textit{Distributed quantum computing} \cite{Crpeau2002,Cuomo2020} for several quantum computers,   allowing to compute tasks as one quantum computer; 
\item[-] \textit{Consensus and agreement tasks} refering to the so-called Byzantine Agreement (problem of decision of group on one output despite the intervention of an adversary). The quantum version \cite{BenOr2005} can reach agreement in $\mathcal{O}(1)$ complexity in comparison with classical complexity $\mathcal{O}\sqrt{n/\log n}$. 
\item[-] \textit{Entangled sensor network} \cite{Proctor2018,Xia2021} allowing improvement in the sensitivity of the sensors and reduction of errors, and evaluating global properties rather than gathering data about specific parts of a system.
\end{itemize}

A quantum network allows direct secure quantum communication between quantum computers, where quantum data can be directly exchanged. This can be useful for effective redistribution of computing tasks according to individual quantum computer performance, mainly when an enormous task can be divided into smaller tasks.
Another case is the quantum cloud, where quantum data can be shared between several quantum computers. Moreover, it is questionable whether it will be possible to build one standalone high-performance quantum computer. The realisation will be more likely via distributed quantum computing \cite{Crpeau2002,Cuomo2020}, where many quantum computers will be connected via the quantum network.

%...................................................................................
\subsubsection{Quantum Key Distribution}\label{sec:QKD}
\begin{itemize}[noitemsep]
\item \textbf{Status: } commercially available (with trusted repeaters)
\item \textbf{Utilisation: } new capability
\item \textbf{Timeline expectation:} short-term
\item \textbf{Main challenges: } secure quantum repeater (quantum memory), security certification of the physical hardware
\end{itemize}

Quantum key distribution (QKD) is the most mature application of quantum communication. The goal is to distribute a secret key between two or more parties for encrypted data distributed via classical channels. Due to the no-cloning theorem, any eavesdropper has to perform a measurement that is detectable by communicating parties. 

The dominant classes of protocols are two: one based on BB84 (Bennett-Brassard 1984) protocol \cite{BB84} and the other the E91 (Ekert 1991) protocol \cite{Ekert1991}. The dominant BB84 protocol is technically simpler but requires a quantum random number generation (see Section~\ref{sec:qrng}), and the providing party has to prepare a key before the distribution. Protocol E91 utilises quantum entanglement that generates the key during the process of distribution, and all parties know the key simultaneously. In this protocol, the quantum random number generator is not required. However, the technical solution with quantum entanglement is more challenging. Both classes of protocols are information-theoretically secure.

Theoretically, the QKD is impenetrable during the transmission. However, the typical vector of attack can focus on the final (receiver/transmitter) or intermediate nodes where the hardware of the software layer can contain vulnerabilities such as bugs in control software, an imperfect single-photon source, parties verification problem, etc. This is an area of very active research. For example, the imperfect physical hardware can be abused by the so-called photon-number-splitting \cite{Brassard2000}, or Trojan-horse \cite{Jain2014} attacks. Here, security certification of the hardware and software is necessary and will take time.

Apart from trusted repeaters, the other weak point is the qubit transfer rate, which is too slow to distribute long keys. A new high transfer rate of single-photon sources can resolve the issue. 

At present, QKD technology is commercially available as a point-point connection at short distances or by using trusted repeaters at large distances. The trusted repeater can be a space satellite, as was demonstrated by China \cite{Yin2017,Yin2017v2}.

%...................................................................................
\subsubsection{Post-Quantum Cryptography}
\begin{itemize}[noitemsep]
\item \textbf{Status: } algorithms ready
\item \textbf{Utilisation impact: } must have
\item \textbf{Timeline expectation:} short-term
\item \textbf{Main challenges: } standardisation, implementation
\end{itemize}

Post-quantum cryptography (sometimes referred to as quantum-proof, quantum-safe or quantum-resistant cryptography) represents an area of encryption techniques that should resist future quantum computer attacks. Presently, this is not true for most of the asymmetric encryption that uses public-key technology.
On the other hand, most of the symmetric cryptographic algorithms and hash functions are considered relatively secure against attacks by quantum computers \cite{Bernstein}. Nevertheless, doubling the symmetric key length is recommended \cite{Bernstein2010}.

Now, several approaches are considered as quantum-resistant. For example, lattice-based cryptography \cite{Hoffstein1998}, supersingular elliptic curve isogeny cryptography \cite{DeFeo2014}, hash-based \cite{Merkle2001} cryptography , multivariate-based \cite{Matsumoto1988} cryptography, code-based cryptography \cite{1978DSNPR..44..114M} and symmetric key quantum resistance.

Unlike QKD, all these algorithms are not provably secure from a mathematical perspective.
Therefore, within the process of standardisation, all these algorithms are rigorously tested and analysed, including the implementation. There is no worst case where a quantum-resistant algorithm with bugs in implementation could be cracked by a classical computer \cite{hackQuantumClassic}.
The most followed standardisation process is the one by the U.S. National Institute of Standards and Technology (NIST). The standardisation process is in the third round \cite{Moody2020}, with three finalists (algorithms based on the lattice, code-based and multivariate) and several alternate candidates. The NIST standardisation process is expected to conclude in 2023-24.
Regardless, more and more commercial vendors are offering new quantum-resistant encryption solutions now.

%...................................................................................
\subsubsection{Quantum Random Number Generator}\label{sec:qrng}
\begin{itemize}[noitemsep]
\item \textbf{Status: } commercially available
\item \textbf{Utilisation impact: } new capability (truly random number generation)
\item \textbf{Timeline expectation:} short-term
\item \textbf{Main challenges: } increasing bit rate
\end{itemize}

Random number generators (RNG) are essential for many applications such as Monte Carlo simulations and integration, crypto operations, statistics and computer games. Nevertheless, the RNG in a classical computer, since it acts deterministically, is not truly random, and is called pseudo-random number generation. However, for many applications, the pseudo-RNG is sufficient. 

On the other hand, generating strong keys is the cornerstone of security, which can be achieved only by truly random RNG. One solution is a quantum random number generator (QRNG) that is hardware-based. 
Moreover, QRNG is a crucial part of BB84-based QKD protocols, to be provably secure.

QRNG can be used for any cryptography and makes all cryptography better. One of the advantages of QRNG is that it can be verified and certificated \cite{ABBOTT2014}, unlike any other RNG.

%---------------------------------------------------------------------------
%---------------------------------------------------------------------------
\subsection{Quantum Sensing and Metrology}

Quantum sensing and metrology is the most mature quantum technology area, which improves timing, sensing or imaging. For example, atomic clocks from the first quantum revolution have been part of the Global Positioning System (GPS) for almost half a century. The current quantum clocks are coming up with much higher time measurement precision.

Quantum sensing stands for all quantum technologies that measure various physical variables such as external magnetic or electric fields, gravity gradient, acceleration and rotation. 
Quantum sensors can produce very precise information about an electric signal, magnetic anomalies and for inertial navigation. 

Quantum imaging is a subfield of quantum optics exploiting photon correlations, allowing suppression of noise and increasing the resolution of the imagined object. 
Quantum imaging protocols are considered for quantum radar, detecting objects in the optically impermeable environment, and in medical imaging.

Quantum sensing and metrology technology relies on one or more of the following features: quantum energy levels, quantum coherence and quantum entanglement \cite{Degen2017}. Individual quantum sensors have various metrics that vary with the application. The common metrics are: sensitivity (a signal that gives unity signal-to-noise ratio after 1 second of integration time), dynamic range (minimal and maximal detectable signal), sampling rate (how often the signal is sampled), operating temperature, etc. Derived key metrics include, for example, spatial resolution at a certain distance and the time required to achieve a specified sensitivity.
The typical measure quantities are magnetic and electric fields, rotation, times, force, temperature and photon counting.

%...................................................................................

\subsubsection{Quantum Electric, Magnetic and Inertial Forces Sensing}
\begin{itemize}[noitemsep]
\item \textbf{Status: } laboratory prototypes
\item \textbf{Utilisation impact: } precision, new capability
\item \textbf{Timeline expectation:} short- to long-term
\item \textbf{Main challenges: } miniaturisation, cooling
\end{itemize}

Many sensing quantum technologies are universal and can measure various physical quantities. A detailed description of each technology is outside the scope of this report; however, a basic overview is provided. Many applications include various quantum technologies. For example, quantum inertial navigation consists of three types of sensing: acceleration, rotation and time. In general, precise quantum-based timing is required for many applications, not only for quantum technologies. For quantum timing, see Section~\ref{sec:clocks}.
The most promising technologies are: atomic vapour, cold-atom interferometry, nitrogen-vacancy centres, superconducting circuits and trapped ions.

\textit{Cold-atom interferometry} (measured quantities: magnetic field, inertial forces, time).
Atoms cooled at very low temperatures exhibit wave-like behaviour and are sensitive to all forces that interact with their mass. The changes are observed in the interference pattern \cite{Degen2017,Barrett2016,UKLandscape2016}. The particular realisation can be in the form of Raman atom interferometry, atom Bloch oscillation or others \cite{PhysRevLett.67.181,PhysRevLett.75.2633,Young2007}.
For example, in gravimetry, the quantum-based gravimeter has the potential to reach about several orders of magnitude higher precision \cite{UKLandscape2016} than the best classical counterparts. Such a precise gravimeter allows very detailed mapping of the Earth's surface and underground with a resolution at the centimetre level.
Regarding inertial navigation, the shaken lattice interferometry has the potential to overcome shortcomings of the state-of-the-art atom interferometry techniques and can work as accelerometer and gyroscope at once \cite{Weidner2018}.
Several challenges remain. Some of the biggest challenges are the integration of the quantum sensor into one quantum inertial measurement unit, miniaturisation of laser cooling apparatus used for cooling down atoms and simultaneously maintaining the coherency (suppression of the interaction with the noisy environment), or the dynamic range of the cold atom sensor outside the laboratory.
However, significant advances also can be found in this area, e.g. \cite{Zhu2020}. For a review see \cite{Geiger2020}.

\textit{Trapped ions} (measured quantities: electric and magnetic field, inertial forces, time).
Trapped ions are one of the most universal sensing platforms \cite{Biercuk2010,Diddams2001,Ivanov2016}. Well-controlled trapped ions form a crystal with quantised modes of motion. Any disturbance can be measured through the transition between these modes. A single trapped ion can serve as a precise measurement of time or as a qubit in a quantum computer.
For inertial navigation, the optical lattice technology of trapped cold atoms in 1, 2 and 3-dimensional arrays potentially offers a sub-cm level in size. 
Besides allowing measurement of gravitational and inertial parameters, it can measure Casimir or van der Waals forces.
More recently, using quantum-entangled trapped ions, measurement of electric fields has reached a sensitivity of $\sim 240$~nV/m$s^{-1}$ \cite{Gilmore2021}, which is several orders of magnitude better than the classical counterpart.

\textit{Nitrogen-vacancy (NV) centres} (measured quantities: electric and magnetic field, rotation, temperature, pressure).
Nitrogen-vacancy centre in a diamond crystal works as an electron spin qubit that couples with external magnetic fields. 
In addition, negatively charged NV centres using Berry's phase can measure rotation. 
In general, NV centre-based sensors offer high sensitivity, cheap production and operation in a wide range of conditions \cite{Degen2017,Taylor2008,Ledbetter2012}. In particular, NV centre-based technology can also work at room temperature and higher.
A novel proposed 3D design allows to sense all three components of magnetism, acceleration, velocity, rotation or gravitation simultaneously \cite{PhysRevLett.116.030801}.
The strengths of NV centres in diamond-based sensing are spatial resolution and sensitivity. On the other hand, the challenge is choosing, implementing and manufacturing individual NV centres or their ensembles. In the case of electric field sensing, it is challenging to define a sensitivity \cite{Radtke2019}.

\textit{Superconducting circuits} (measured quantities: electric and magnetic field).
The technology of superconducting circuits based on the Josephson effect describes the quantum tunnelling effect between two superconductors \cite{Degen2017}. This technology allows manufacturing a quantum system at the macroscopic scale and can be controlled effectively with microwave signals.
The superconducting quantum interference device (SQUID) is one of the best magnetometric sensors. However, the disadvantage is the requirement of cryogenics. 
Note that for the measurement of magnetic-field variations smaller than the geomagnetic noise, the preferred design is based on an array of sensors to cancel the spatial-correlation with applications, such as in medical and biomedical applications (e.g. MRI or molecule tagging).
The recent development shows that the superconducting qubits used in quantum computers can be used to measure electric and magnetic fields \cite{Degen2017} as well.

\textit{Atomic vapour} (measured quantities: magnetic field, rotation, time). Spin-polarised high-density atomic vapour undergoes state transition under external magnetic field which can be measured optically \cite{Degen2017,Dang2010,PhysRevLett.104.133601}.
An advantage is a deployment at room temperature.
The atomic vapour is suitable for rotation sensing, known as the Atomic Spin Gyroscopes (AGS). AGS can be chip-scale \cite{UKLandscape2016}.    
For comparison, the best classical rotation sensors are very precise (e.g. ring laser gyroscope). The expected quantum sensor will be about twice as precise. However, the mentioned best classical gyroscope has a size of $4\times4$~m, which is impractical \cite{Schreiber2013}.    
Atomic vapour cell magnetometers based on atomic ensembles have the potential to outperform SQUID magnetometers and work at room temperature \cite{Degen2017}.

%...................................................................................
\subsubsection{Quantum Clocks}\label{sec:clocks}
\begin{itemize}[noitemsep]
\item \textbf{Status: } laboratory prototypes
\item \textbf{Utilisation impact: } precision
\item \textbf{Timeline expectation:} short- to mid-term
\item \textbf{Main challenges: } miniaturisation
\end{itemize}

Atomic clocks have been with us for several decades; for example, as part of GPS satellites. The current atomic clocks are based on atomic physics, where the electromagnetic emissions from electrons when changing energy level utilise a `tick'. As such, the atomic clock is a very mature technology. Today, the atomic clocks based on atomic fountain or thermal atomic beam and magnetic state selection principles can reach a relative uncertainty $\sim 10^{-15}- 10^{-16}$ \cite{BIPM}, or state-of-the-art chip-size atomic clocks have uncertainty $2\times 10^{-12}$ \cite{UKLandscape2016}.

The second quantum revolution comes with new principles for atomic or quantum clocks.
Quantum logic clock is based on single-ion, a technology related to trapped-ion qubit for quantum computing \cite{Diddams2001}.
Quantum logic clock was the first with clock uncertainty below $10^{-18}$ \cite{Brewer2019}. Quantum clocks can also benefit from quantum entanglement \cite{Hwang2002}.

Later, the quantum logic clock was superseded by experimental optical lattice clocks. Note that the current atomic clocks work with microwave frequencies, i.e. the transition between energy levels emits a microwave photon. The measurement of level transition with the emitted photon in optical frequencies is harder to achieve, although it offers better performance. Optical clocks are still in development, with systems being based on: single ions isolated in an ion trap, neutral atoms trapped in an optical lattice and atoms packed in a 3D quantum gas optical lattice. The 3D quantum gas optical lattice clocks in particular have demonstrated frequency precision $2.5\times 10^{-19}$ \cite{Marti2018}. Recently, it was demonstrated that quantum entanglement could enhance the clock stability \cite{PedrozoPeafiel2020}.

Another research focuses on vapour-cell (or gas-cell) atomic clock that provides chip-size realisation \cite{Camparo2007}; 
solid-state (for instance, the NV centre in diamond) clock \cite{Hodges2013}; or
nuclear clock with a similar principle as microwave or optical atomic clock, except that it uses nuclear transition instead of electron transition in an atom's shell \cite{vonderWense2020}, with the potential for unprecedented performance, outstripping atomic optical clocks \cite{Campbell2012}.

Various clock technologies have their own challenges, such as precise frequency comb, laser system for control and cooling down and black body radiation shift (in the case of optical clocks).
Also, miniaturisation usually comes at the cost of lower frequency precision. Another common type of challenge is the synchronisation of those clocks.

Precise timing is essential for many technologies, such as satellite navigation, space systems, precise measurement, telecommunication, defence, network synchronisation, finance industry, energy grid control, and in almost all industrial control systems. 
However, very precise timing is crucial for quantum technologies, especially for quantum sensing and imaging.
For instance, a very high precision clock allows new measurements, such as gravity potential measurement down to the centimetre level at the Earth's surface or searching for new physics.

%...................................................................................
\subsubsection{Quantum RF Antenna}
\begin{itemize}[noitemsep]
\item \textbf{Status: } laboratory prototypes
\item \textbf{Utilisation: } effectiveness
\item \textbf{Timeline expectation:} short- to mid-term
\item \textbf{Main challenges: } miniaturisation, cooling
\end{itemize}

Radio frequency (RF) antennas serve as receivers or transmitters of various signals. They can be simple dipole antennas to complex AESA\footnote{AESA (active electronically scanned array) module consists of an array of small transmitters and receivers that are controlled by a computer. Simply put, each cell of AESA can behave as an independent radar module.} modules. Their size limitation is bounded by the wavelength of the produced or received signal. For example, a 3~GHz signal has a wavelength of $\sim 10$~cm and the size of the antenna should be no less than approximately $1/3$ of this wavelength. This is called the Chu–Harrington limit \cite{Chu1948,Harrington1960}. 

Rydberg atoms' technology allows breaking this limit and having an antenna of the size of a few micrometres independently on the receiving signal wavelength. Rydberg atoms are highly excited atoms with a correspondingly large electric dipole moment, and therefore high sensitivity to external electric field \cite{Facon2016,Cox2018}. 
Note that Rydberg atoms-based antenna can only receive a signal.

The recent prototypes of Rydberg atoms-based analyser were demonstrated for frequencies 0 to 20~GHz for AM or FM radio, WiFi and Bluetooth signals \cite{Meyer2021}. The combination of more antennas can detect the angle-of-arrival of the signal \cite{robinson2021determining}.
At the laboratory level, Rydberg atoms technology is available commercially.

Quantum RF receiver as a single cell (for targeted frequency, narrow bandwidth) or arrayed sensor (broad frequency span) can find its applications in navigation, active imaging (radar), telecommunications, media receiver or passive THz imaging.

%...................................................................................
\subsubsection{Quantum Imaging Systems}\label{sec:qimag}
\begin{itemize}[noitemsep]
\item \textbf{Status: } laboratory prototypes and proof-of-concept verifications
\item \textbf{Utilisation: } new capability
\item \textbf{Timeline expectation:} short- to long term
\item \textbf{Main challenges: } improving resolution, high rate single-photon sources
\end{itemize}

Quantum imaging systems are a wide area covering 3D quantum cameras, behind-the-corner cameras, low brightness imaging and quantum radar or lidar (for quantum radar, see Section~\ref{sec:QTquantumradar}).

\textit{SPAD (Single Photon Avalanche Detectors) Array} is a very sensitive single-photon detector connected with a pulsed illumination source that can measure the time-of-flight from source to an object and hence the range of the object. Then, putting SPAD into an array can work as a 3D camera. SPAD works with the optical spectrum with extension developed to the near-infrared spectrum.

SPAD array can be used to detect objects out of the line of sight, too (e.g. hidden behind the corner of a wall). The idea is based on laser and camera cooperation, where the laser sends a pulse in front (e.g. a spot on the floor) of the SPAD camera. From the spot, the laser pulse will scatter in all directions, including behind the corner, where the photons can be reflected to the spot in front of the SPAD camera and then to the camera. SPAD is sensitive enough to detect such a three-scattered signal \cite{Gariepy2015}.

\textit{Quantum ghost imaging} \cite{Aspden2015,Moreau2017,Meyers2015}, also known as coincidence imaging or two-photon imaging, is a technique that allows imaging an object that is out of the line of sight of the camera.
In the source, two entangled photons are created, each of a different frequency. The one in the optical frequency is recorded directly by a high-resolution photon-counting camera.
The second photon having a different frequency (e.g. the infrared) is sent toward the object. The reflected photon is detected by a single-photon detector (the so-called `bucket' detector). The image is then created from the correlations between both photons. The ghost imaging protocol was demonstrated without quantum entanglement, too (using classical correlation), although with worse resolution. 

Such a schema allows imaging an object at extremely low light levels. Also, infrared light can better penetrate some environments with a better signal-to-noise ratio (SNR) \cite{Walborn2010}. Ghost imaging experiments that use x-ray or ultra-relativistic electrons were demonstrated recently \cite{Pelliccia2016,Li2018}.

\textit{Sub-shot-noise imaging} \cite{Brida2009} is another quantum optics schema allowing detection of a weak absorption object with a signal below the shot noise. Shot noise is the result of fluctuations in the detected number of photons. For example, the shot noise is the limit for lasers. 
This limit can be overcome using correlated photons. The detection of one `herald' or `ancilla' photon signifies the presence of the correlated photon that probes the object or environment.

\textit{Quantum Illumination} (QI) \cite{Lloyd2008} is a quantum protocol to detect a target using two correlated (entangled) photons. One photon, called the `idler', is kept. The other, called the `signal' photon, is sent toward the target and reflected, and both photons are measured. The advantage of this protocol remains even when the entanglement is destroyed by a lossy and noisy environment. QI protocol is one of those mainly adapted for the quantum radar, but it can also be applied to medical imaging or quantum communication.

%...................................................................................
\subsubsection{Quantum Radar Technology}\label{sec:QTquantumradar}
\begin{itemize}[noitemsep]
\item \textbf{Status: } laboratory prototypes and proof-of-concept verification
\item \textbf{Utilisation: } new capability
\item \textbf{Timeline expectation:} long-term and more
\item \textbf{Main challenges: } high rate single-photon source, quantum microwave technology
\end{itemize}

Quantum radar, in principle, works similarly to classical radar, in the sense that a signal has to be sent toward the target, and the radar system needs to wait for the reflected signal. Nevertheless, theoretically improved precision and new capabilities can be achieved by quantum mechanical approaching.

There are several protocols considered for quantum radar, such as interferometric quantum radar \cite{SmithIII2009}, quantum illumination (QI) \cite{Lloyd2008}, hybrid quantum radar \cite{Barzanjeh2020,Luong2020} or Maccone-Ren quantum radar \cite{Maccone2020}. None of the mentioned protocols is perfect. Interferometric quantum radar, for example, is too sensitive to noise and requires quantum entanglement preservation. QI is an ideal protocol for a noisy environment and is even laboratory-verified for microwave spectrum \cite{Barzanjeh2015}, but it requires knowledge of the distance to the target, and such as it has no ranging function. Nevertheless, the QI-based approach to quantum target ranging is under development \cite{karsa2021energetic}.
This ranging problem is also solved by the hybrid quantum radar, but at the expense of sensitivity. The Maccone-Ren protocol has QI properties and ranging function, but it is only a theoretical concept so far.

The biggest challenge common to all protocols is the high rate of generation of entangled photons in (not only) a microwave regime. The quantum version of the radar equation \cite{lanzagorta2011quantum} still holds the dominant term $1/R^4$, where $R$ is the radar--target distance.
As a result, the number of demanded entangled photons (modes) is several orders of magnitude higher than is available currently \cite{Pirandola2018}. In a sense, quantum radar is similar to noise radars and shares many properties such as the probability of interception, low probability of detection, efficient spectrum sharing, etc., see \cite{Luong2020} and references therein.

Another related challenge is target finding. Theoretical work \cite{Zhuang2020} shows that quantum entanglement can outperform any classical strategy in finding the unknown position of the target. Moreover, the presented method can work as a quantum-enhanced frequency scanner for the fixed target range. 

%...................................................................................
\subsubsection{Other Sensors and Technology}
\begin{itemize}[noitemsep]
\item \textbf{Status: } laboratory prototypes
\item \textbf{Utilisation: } new capability (e.g. chemical and precise acoustic detection)
\item \textbf{Timeline expectation:} short- to medium term
\item \textbf{Main challenges: } improving resolution
\end{itemize}

Quantum technologies can be used for ultra-precise sound sensing up to the level of a phonon, a quasiparticle quantising sound waves in solid matter \cite{Chu2017,Satzinger2018}, using photoacoustic detection.
Precise detection of acoustic waves is essential for many applications, including medical diagnostics, sonar, navigation, trace gas sensing and industrial processes \cite{Wu2017,Fischer2016}.

Photoacoustic detection can be combined with quantum cascade laser and used for gas or general chemical detection. Quantum cascade laser (QCL) is yet a mature technology \cite{Faist1994}.  
QCL is a semiconductor laser that emits in the mid- and long-wave IR bands and, as with many other quantum technologies, requires cooling far below $-70^\circ$C.
However, recent development allows chip-level implementation working at around $-23^\circ$C, which can be achieved by a portable cooling system \cite{Khalatpour2020}.

%---------------------------------------------------------------------------
%---------------------------------------------------------------------------
%---------------------------------------------------------------------------
\section{Quantum Technology In Defence}\label{sec:strategy}

Military technologies have more demanding requirements than industrial or public applications.
This requires greater caution, considering possible deployment on the battlefield.
Sec.~\ref{sec:warfare} presents various possible military applications with different TRLs, time expectations and with multiple risks of realisation.

It will be simpler and less risky for technologies that are easily implemented and fit into current technologies, such as quantum sensors where, simply put, we can replace a classical sensor with a quantum sensor. 

On the contrary, QKD is an example of a technology that is already commercially available but is challenging to deploy. A lot of new hardware, systems and interoperability with current communication systems are needed. Thus, this technology carries more significant risks in terms of military deployment.

We can expect an advantage in lowering SWaP and scaling up quantum computers and quantum networks in the long term. That will make the deployment easier and probably necessary if the nation/army wants to compete with other nations/armies with edge (quantum) technologies.

%---------------------------------------------------------------------------
\subsection{Quantum Strategy}

The future users of military quantum technologies will have to think carefully about whether, where and when to invest time and resources.
The goal of the defence forces is not to develop military technology but usually only to specify requirements and their acquisition. However, they can participate significantly in development, especially if they are the end user.

As a foundation, it is ideal to have a national quantum ecosystem in place composed of industry and academic institutions.
Such an ecosystem should be supported generally at the government level, i.e. having a national quantum plan, but should also be motivated to develop technologies for the defence sector.
This can be achieved through appropriate grant funding and even various thematic challenges, in which individuals and startups can participate and perhaps bring new disruptive ideas and solutions.
This will naturally lead to closer cooperation with industry and academia. The quantum industry is quite interesting, where there is a great deal of cooperation between academia and industry.

The first step is to establish a quantum technology roadmap or quantum strategy. The roadmap/strategy should specify all the next steps, from identifying disruptive quantum solutions, market survey, technology and risk assessment and development itself to prototype testing and eventually solution deployment. The roadmap or quantum strategy can consist of three parts: 
\begin{enumerate}[noitemsep]
  \item Identification,
  \item Development,
  \item Implementation and deployment.
\end{enumerate}

The most critical part is the identification of the most advantageous and disruptive quantum technologies for the considered warfare domains. This step also includes the technological and scientific assessment to balance technological risk (limited deployability, performance below expectations, or impossibility of transfer from the laboratory to the battlefield) versus the potential advantage of individual quantum technologies.
This process of identification should be repeated in cycles in order to react relatively quickly to new discoveries and disruptive solutions. 
It is important to remember that many applications are yet to be identified or discovered.

The next step is the usual process of research and development (R\&D). The R\&D should be sufficiently supported financially, but also with minimal bureaucratic obstacles. It should involve fast development cycles with close interaction with the end user of the military technology (specifications and performance consultations, prototype testing, preparing for certifications, ...). At the end of this phase, the new system at the initial operating capability should be ready.

The last step is to reach full operational capability, including modification or creation of new military doctrines, preparing new military scenarios, strategies and tactics fully exploiting the quantum advantage.

The final note pertains to the Identification phase.
Here, the decision maker needs to also assume the long-term perspective. So far, many quantum technologies have been considered individually: sensors, QKD, quantum computing, etc. 
However, the long-term vision considers the interconnection of quantum sensors and quantum computing via the quantum network. Here, the theoretical and experimental works demonstrate additional quantum advantage exploiting quantum entangled sensors and computers \cite{Proctor2018,Xia2021}.
More similar applications may yet be discovered or invented.
This is important to consider when the optical-fibre/quantum networks are being built. Later, the current elements such as trusted repeater can be replaced by fully quantum repeaters and switches, allowing to reach the full potential of the quantum network.

%---------------------------------------------------------------------------
\subsection{TRL and Time Horizon}

As has been mentioned several times, various quantum technologies are at different TRL, varying from 1 to 8.
The TRL variation and time horizon expectations are even more complex when considering various applications and deployment platforms, especially for military purposes.
Some TRL and time horizon estimates were provided in \cite{NATOTrends}.
However, some estimations, such as quantum precision navigation at TRL 6, seem too optimistic based on what is described in this report. 

Here, we provide our own TRL and expected time horizon in Tab.~\ref{tab:TRLhor}, which correspond to the findings of this work.
\begin{table}[!htb]
    \centering
    \begin{tabular}{l|c|c}
        \hline
        \textbf{Technology} & \textbf{TRL} & \textbf{Horizon} \\
        \hline
        \hline
         Quantum computer (annealer) & 4-5 (5-6) & 2030 \\
         QKD (satellite) & 7-8 (6-7) & 2025 (2030) \\
         Post-quantum cryptography & 7-8 & 2025 \\
         Quantum communication network & 1-3 & 2030-2035 \\
         Quantum inertial navigation & 4-5 & 2025-2030 \\
         Quantum clocks & 4-6 & 2030 \\
         Quantum radar & 1-2 & none \\
         Quantum RF antenna & 4 & 2025-2030 \\
         Quantum magnetic and gravity sensing & 5-6 & 2025 \\
         Quantum imaging & 5 & 2025-2030 \\
         \hline
    \end{tabular}
    \caption{TRL and time horizon expectations. These expectation reflect general TRL rather than just military TRL. Note that various quantum technologies are at different TRL within the same application.}
    \label{tab:TRLhor}
\end{table} 

The reader can compare these with other timelines in \cite{AusQRoadmap,NATOTrends}.

The actual military deployment can take some time to overcome all technological obstacles and meet military requirements.
Take, for example, the quantum gravimeter for undeground scanning.
The first generation will likely be deployed as a static sensor placed on a truck, and the range/spatial resolution will be rather low.
In time, the next generation will improve sensitivity and spatial resolution. Along with reduction of SWaP, the sensor will be capable of being placed aboard an aircraft, and later on a drone and maybe on an LEO satellite. However, it is also possible that the sensor's limits will be reached earlier, resulting in  deployment becoming impossible, e.g. on a drone or LEO satellite.

%---------------------------------------------------------------------------
\subsection{Quantum Technology Countermeasures}

A standalone section on quantum technology countermeasures is warranted, although this topic will be touched upon, e.g. in Sec.~\ref{sec:quantumEW} about the quantum analogy of the classical electronic warfare. 
This topic is less studied, and few texts deal with this subject; besides, a detailed description is beyond the scope of this report.

Briefly, this topic refers to the methods and techniques of spoofing, disabling or destroying quantum technologies, whether it is quantum computers, quantum networks or quantum sensors and imaging systems. 
Quantum technologies exploit the quantum-physical properties of individual quanta. As such, they are very susceptible to interference and noise from the environment, and so can potentially be spoofed or paralysed.
Especially in relation to quantum networks and in particular to QKD, we speak about quantum hacking \cite{Makarov2005,Zhao2008,Lydersen2010,Gerhardt2011,Bugge2014}, which has developed hand in hand with QKD itself.  

Authors and decision makers on quantum strategy should keep in mind that when quantum technologies are deployed in the military field, various countermeasures will very likely emerge sooner or later.
What is currently unknown is the possible effectiveness of quantum technology countermeasures and their impact.

%---------------------------------------------------------------------------
%---------------------------------------------------------------------------
%---------------------------------------------------------------------------
\section{Quantum Technology Military Applications}\label{sec:warfare}

\begin{figure}[htb]
    \centering
\includegraphics[width=\textwidth]{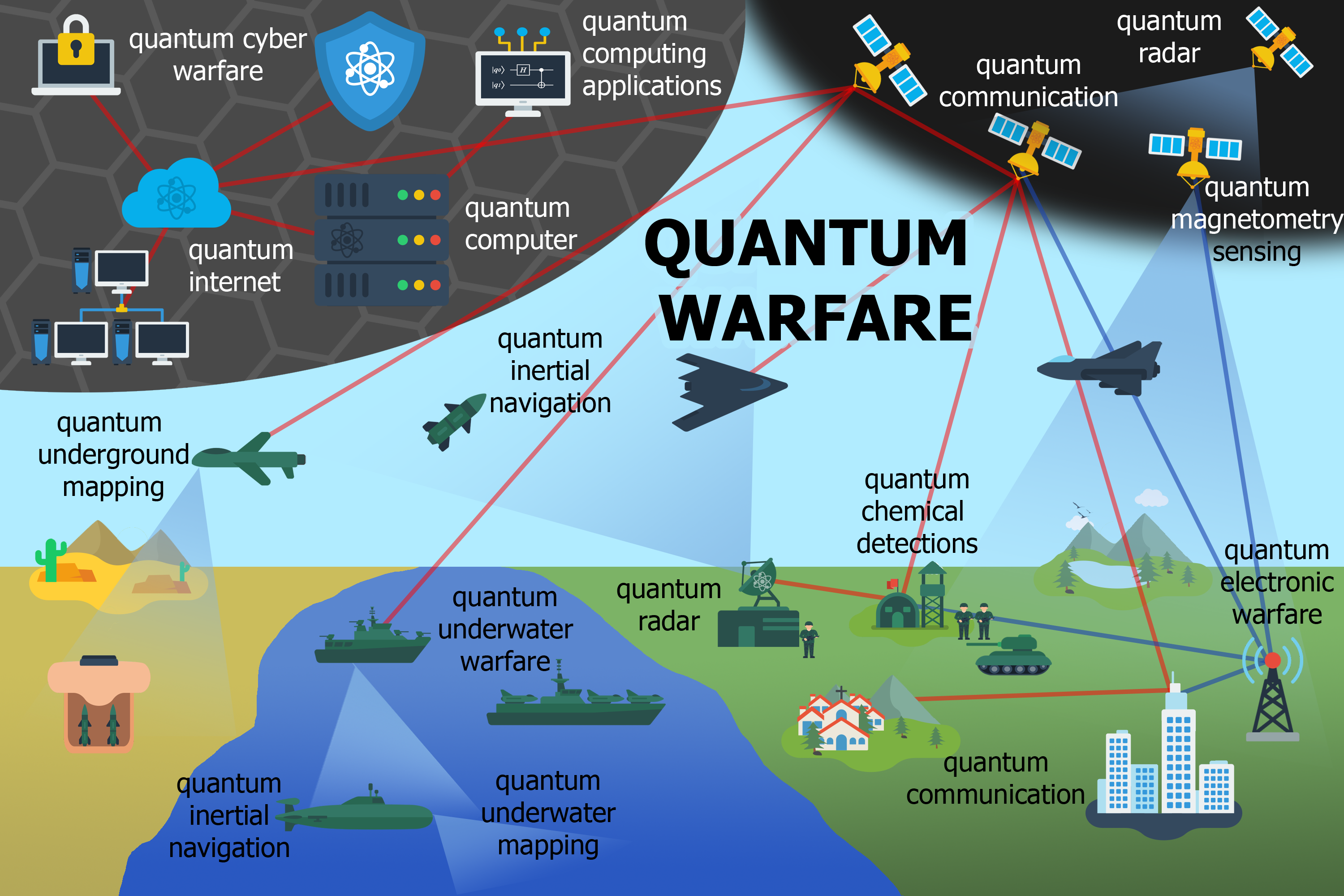}
    \caption{Sketch of quantum warfare utilising various quantum technology systems.}
    \label{fig:qw}
\end{figure}

Quantum technologies have the potential to significantly affect many areas of human activity. This is especially true for the defence sector. Quantum technologies can impact all the domains of modern warfare. The second quantum revolution will improve sensitivity and efficiency, and introduce new capabilities and sharpen modern warfare techniques rather than lead to new types of weapons.

The following text maps the conceivable quantum technology applications for military, security, space and intelligence in different aspects of modern warfare. It also mentions the industrial applications which may suggest quantum technologies' capabilities and performances, especially when no public information on military applications is available.

It is important to notice that many applications are still more theoretical than realistic. The significant quantum advancement achieved in the laboratory does not always result in similar progress outside the laboratory. 
The transfer from laboratory to practical deployment involves other aspects too, such as portability, sensitivity, resolution, speed, robustness, low SWaP (size, weight and power) and cost, apart from a working laboratory prototype.
The practicality and cost-effectiveness of quantum technologies will determine whether particular quantum technologies are manufactured and deployed.

The integration of quantum technology into a military platform is even more challenging. Apart from quantum computers that will mostly be located at data centres similarly as for civil use, the integration and deployment of quantum sensing, imaging and networks faces several challenges posed by the increased demands of military use (in comparison with civil/industry or scientific requirements). For example, the military level requirement of precise navigation necessitates fast measurement rates that can be quite limiting for the current quantum inertial sensors. There are more examples, and probably more are yet to come.

Moreover, this area is still very young, and new technological surprises, both in a bad and a good sense, could impose other quantum advantages or disadvantages.

%...................................................................................
\subsection{Quantum Cybersecurity}

Key points:
\begin{itemize}[noitemsep]
\item Necessity of quantum crypto-agility implementation.
\item Operations that want to take advantage of Shor's algorithm should start to collect the data of interest before the quantum-safe encryption is deployed.
\item The implementation of QKD needs to be carefully considered.
\item In QKD, the endpoints will be the weakest part of the system. 
\end{itemize}

Quantum advantage in cyber warfare can provide new, but on the one hand very effective (with exponential speedup), vectors of attack on the current asymmetric encryptions (based on integer factorisation, the discrete logarithm or the elliptic-curve discrete logarithm problem) and, theoretically, on symmetric encryption \cite{RR-3102-RC,Moody2020}.
On the other side are new quantum-resilient encryption algorithms and approaches, as well as quantum key distribution. 
For an overview, see, for example, \cite{Pirandola2020,Lindsay2020,Hudson2018,EC2018}.

The current trend also is the development and employment of machine learning or artificial intelligence for cyber warfare \cite{Kline2019}. For more details on the quantum opportunities, see Section~\ref{sec:QCcap}.

\subsubsection{Quantum Defence Capabilities}

The post-quantum cryptography implementation is the `must-have' technology that should be carried out as soon as possible. The risk that hostile intelligence is gathering encrypted data with the expectation of future decryption using the power of quantum computers is real, high and present \cite{Thales2020}. This applies to military, intelligence and government sectors as well as to industry or academia where secrets and confidential data are exchanged or stored.
The current trend is to start preparing the infrastructure for implementing quantum crypto-agility when the certified (standardised) post-quantum cryptography becomes ready to deploy \cite{RR-3102-RC,Moody2020}.  

New quantum-resilient algorithms can offer not only a new mathematical approach difficult enough even for quantum computers, but also a new paradigm of working with encrypted data. For instance, fully homomorphic encryption (FHE) allows the data to never get decrypted---even if they are being processed \cite{FHE}. Although the security applications, such as for genomic data, medical records or financial information, are the most mentioned, applications for intelligence, military or government are evident, too. As such, FHE is a good candidate for cloud-based quantum computing to ensure secure cloud quantum computation \cite{Huang2016}. 

Note that post-quantum cryptography should be implemented in the Internet of Things (IoT), or the Internet of Military Things (IoMT) \cite{IoMT}, as a rapidly growing sector with many potential security breaches. For an overview of post-quantum cryptography for IoT, see \cite{FernandezCarames2020}.

Quantum key distribution (QKD) \cite{moskovich2015overview,ETSI2018,EC2018} is another new capability that allows safe encryption key exchange where the security is mathematically proven. Although it is impossible to eavesdrop on the quantum carrier of the quantum data (key), the weaknesses can be found at the end nodes and trusted repeaters, due to imperfect hardware or software implementation. 
Another question is the cost, considering the quantum data throughput, security and non-quantum alternatives independently if the solution is optical fibre-based or utilising quantum satellites. The QKD solution seems to be preferred in EU \cite{EuroQCI}, while the post-quantum encryption solution finds favour in US \cite{NAS-QC}. 

The last note refers to quantum random number generators. QRNG increases security \cite{Abellan2018} and denies attacks on pseudorandom number generators \cite{10.1007/3-540-69710-1_12}.

\subsubsection{Quantum Attack Capabilities}

With Shor's algorithm-based quantum cryptoanalysis of Public key encryption (PKE)---for instance, RSA, DH, ECC---the attacker can decrypt the encrypted data collected earlier. There is no precise forecast when the so-called `Q-Day', the day when a quantum computer breaks the 2048-bit RSA encryption, will happen. However, the general opinion is it will take about 10--15 years (based on a survey in 2017) \cite{QuantumRisk2017}. A similar threat applies to most message authentication codes (MAC) and authenticated encryption with associated data (AEAD), such as HMAC-CBC and AES-GCM, because of Simon's algorithm and superposition queries.

One has to assume that such offensive operations already exist or that intense research is being done.
In 10 years, most sensitive communication or subjects of interest will be using the post-quantum cryptography or QKD implemented in the next six years. That means by the time a quantum computer able to crack PKE becomes available, most of the security-sensitive data will be using a quantum-safe solution.

In theory, Grover's algorithm weakens the symmetric key encryption algorithms; for example, DES and AES.
However, the quantum computing, and in particular quantum memory, requirements are so huge that it seems to be unfeasible in the next few decades \cite{cryptoeprint:2019:1208}.

Another vector of attack uses the classical hacking methods of classical computers that will remain behind quantum technologies. In general, quantum technology is a technologically young sector where plenty of new quantum system control software is being developed. The new software and the hardware tend to have more bugs and security breaches.
For example, the current QKD quantum satellites working as trusted repeaters controlled by a classical computer can be an ideal target for a cyber attack. 
Moreover, specific physical-based vectors of attack against quantum networks (e.g. QKD) are the subject of active research \cite{satoh2020attacking}, such as photon-number-splitting \cite{Brassard2000} or the Trojan-horse attack \cite{Jain2014}, and future surprises cannot be excluded.
For an overview of quantum hacking, see, e.g. \cite{Pirandola2020}.

%...................................................................................
\subsection{Quantum Computing Capabilities}\label{sec:QCcap}

Key points:
\begin{itemize}[noitemsep]
\item Quantum computing capabilities will increase with the number of logical qubits.
\item Most likely, quantum computing will be used as part of a hybrid cloud.
\item Small, embedded quantum computing systems are desirable for direct quantum data processing.
\item General use for quantum optimisations, ML/AI enhancements and faster numerical simulations.
\end{itemize}

Quantum computing will introduce new capabilities to the current classical computing services, helping with computational problems of high complexity. Further, besides the quantum simulations described above, quantum computing covers quantum optimisations, machine learning and artificial intelligence (ML/AI) improvement, quantum data analysis, and faster numerical modelling \cite{national2019quantum,AusQRoadmap}.
The military problems that could be solved with near-term quantum computers were presented in \cite{ATARC}. They are: Battlefield or war simulations; Analysis of radio frequency spectrum; Logistics management; Supply chain optimisation; Energy management; and Predictive maintenance.

To get the  most effective results, future quantum computing implementation will be in computing farms along with classical computers, which will create a hybrid system. A hybrid quantum-classical operating system will analyse the tasks to be computed using ML/AI, and split individual computations into resources such as CPU, GPU, FPGA\footnote{Central Processing Unit; Graphics Processing Unit; Field-Programmable Gate Array}, or quantum processor (QPU), where the best and fastest result can be obtained. 

A small, embedded quantum computer that could be placed, for example, in an autonomous vehicle or mobile command centre is questionable. The current most advanced qubit designs need cryogenic cooling. Therefore, more efforts should be focused on the other qubit designs as photonic, spin or NV centres that can work at room temperature.
The embedded quantum chip could perform simple analytical tasks or serve for simple operations related to quantum network applications where a straightforward quantum data  process is desired.
Nevertheless, the machine learning and model optimisation of autonomous systems and robotics can also benefit from `large' quantum computers.

Quantum computing is likely to be efficient in optimisation problems \cite{Lavoix2019,Uppal2020,ATARC}. In the military sector, examples of quantum optimisations could be logistics for overseas operations and deployment, mission planning, war games, systems validation and verification, new vehicles' design and their attributes such as stealth or agility. At the top will be an application for enhanced decision making, supporting military operations and functions through quantum information science, including predictive analytics and ML/AI \cite{Wilson2020}.
Specifically, quantum annealers have proven themselves in verifying and validating complex systems' software code \cite{Allen2010,LM2017}.

Quantum computers are expected to play a significant role in Command and Control (C2) systems. The role of C2 systems is to analyse and present situational awareness or assist with planning and monitoring, including simulation of various possible scenarios to provide the best conditions for the best decision. Quantum computers can improve and speed up the scenario simulations or process and analyse the Big Data from ISR (Intelligence, Surveillance and Reconnaissance) for enhanced situational awareness. This also includes the involvement of quantum-enhanced machine learning and quantum sensors and imaging.

Quantum information processing will probably be essential for Intelligence, Surveillance, and Reconnaissance(ISR) or situational awareness. ISR will benefit from quantum computing, which offers a considerable boost to the ability to filter, decode, correlate and identify features in signals and images captured by ISR.
Quantum image processing in particular is an area of extensive interest and development. It is expected that in the near term situational awareness and understanding can benefit from quantum image analysis and pattern detection utilising neural networks \cite{DSTL-QC}.

Quantum computing will enhance classical machine learning and artificial intelligence \cite{Biamonte2017}, including for defence applications \cite{Wilson2020}. Here, quantum computing will surely not be practical to carry out the complete machine learning process. Nevertheless, quantum computing can improve ML/AI machinery (e.g. quantum sampling, linear algebra, quantum neural networks). A recent study \cite{Huang2021} shows that quantum ML provides an advantage just for some kernels fitting particular problems.
Quantum computing can possibly enhance, in principle, most classical ML/AI applications in defence; for example, automating cyber operations, algorithmic targeting, situation awareness and understanding and automated mission planning \cite{AIdef2019,Spiegeleire2017}. 
The most immediate application of quantum ML/AI is probably quantum data; for instance, data produced by quantum sensing or measuring apparatus \cite{Dunjko2016}.
Actual applicability will grow with quantum computer resources, and in eight years, quantum ML/AI can be one of the important quantum computing applications \cite{WitterInterview}. Such applicability can be accelerated by hybrid classical-quantum machine learning where tensor network models could be implemented on small near-term quantum devices \cite{Huggins2019}.

Quantum computers, through quantum neural networks, can be expected to provide superior pattern recognition and higher speed. This may be essential, for instance, in bio-mimetic cyber defence systems that protect networks, analogously to the immune systems of biological organisms \cite{DSTL-QC}.

Besides, through faster linear algebra (see \ref{sec:linalg}), quantum computing has the potential to improve the current numerical linear equation-based numerical modelling in the defence sector, such as war games simulations, radar cross section calculations, stealth design modelling, etc. 

In the long term, the quantum systems can enable Network Quantum Enabled Capability (NQEC) \cite{DSTL-QC}.
NQEC is a futuristic system that allows communication and sharing information across the network between individual units and the commander to respond quickly to battlefield developments and for coordination. Quantum enhancement can bring secured communication, enhanced situational awareness and understanding, remote quantum sensor output fusing and processing, and improved C2.

%One of the few currently known use cases of quantum computing applications is by Lockheed Martin. Lockheed Martin uses quantum annealer from D-Wave Systems to verify and validate complex systems' software code \cite{Allen2010,LM2017}. 

%...................................................................................
\subsection{Quantum Communication Network}
 
Key points:
\begin{itemize}[noitemsep]
\item Various security applications (e.g. QKD, identification and authentication, digital signatures).
\item The adoption of security applications will happen as quickly as all new technology security aspects are explored, carefully.
\item Quantum clock synchronisation allows utilising higher precision quantum clocks.
\item Quantum internet is the most effective way of communication between quantum computers and/or quantum clouds.
\end{itemize} 
 
Quantum internet stands for a quantum network with various services \cite{neumann2020quantum} which have significant, and not only security, implications. 
However, many progressive quantum communication network applications require quantum entanglement; that is, they require quantum repeater and quantum switch. Recall that the trusted repeaters can be used for QKD only (see Sec.~\ref{sec:qunet}). Future combinations of optical fibre and free-space channels will interconnect various end nodes such as drones, planes, ships, vehicles, soldiers, command centres, etc.

\subsubsection{Security Applications}

Quantum key distribution is one of the most matured quantum network applications. This technology is going to be interesting for the defence sector later, when long-distance communication using MDI-QKD or quantum repeaters becomes possible. Currently, basic commercial technology that uses trusted repeaters is available. These pioneers can serve as a model of how quantum technologies can be employed. Here, QKD companies promote the technology as the most secure, and more and more use cases appear, especially in the financial and healthcare sectors. On the other hand, the numerous recommendation reports and authorities are more circumspect; for example, the UK National Cyber Security Centre \cite{NCSC-QKD} that does not endorse QKD for any government or military applications in its current state.

Apart from QKD, which distributes the key only, the quantum network could be used for quantum-secure direct communication (QSDC) \cite{Long2002,Bostrm2002,Zhang2017,Qi2019} between space, special forces, air, navy and land assets. 
Here, the direct messages encrypted in quantum data take advantage of security similar to QKD.
One obstacle could be a low qubit rate, which will only allow sending simple messages and not audiovisual and complex telemetry data. 
In that case, the network switch to the QKD protocol for distributing the key and the encrypted data will be distributed over classical channels.
Other protocols such as quantum dialogue \cite{Gao2010} and quantum direct secret sharing \cite{Han2008} aim to use the quantum network for provable secure communications as QSDC.
Note that QKD and QSDC are considered to be a native part of 6G wireless communication networks and discussed accordingly in \cite{You2020}.

Another significant contribution of the quantum approach to security is the quantum digital signature (QDS) \cite{Gottesman2001}. It is the quantum mechanical equivalent of a classical digital signature. QDS provides security against tampering of a message after a sender has signed the message.

Next, quantum secure identification exploits quantum features allowing identification without revealing authentication credentials \cite{Damgrd2014}. Non-quantum identification is based on the exchange of login and password or cryptographic keys, which allows intruders to at least guess who has tried to authenticate.  

The other application is position-based quantum cryptography \cite{Malaney2010,Chakraborty2015}.
Position-based quantum cryptography can offer more secure communication, where the accessed information will be available only from a particular geographical position, such as communication with military satellites only from particular military bases. Position-based quantum cryptography can also provide secure communication when the geographical position of a party is its only credential.

\subsubsection{Technical Applications}

Quantum network will perform network clock synchronisation \cite{Chuang2000,Kmr2014} that is already a major topic in classical digital networks. Clock synchronisation aims to coordinate otherwise independent clocks, especially atomic clocks (e.g. in GPS) and local digital clocks (e.g. in digital computers). A quantum network that uses quantum entanglement will reach even more accurate synchronisation, especially when quantum clocks come to be deployed (for Time standards and frequency transfer see Sec.~\ref{sec:PNT}). Otherwise, the high precision of quantum clocks would be utilised locally only. 
Precise clock synchronisation is essential for the cooperation of C4ISR (Command, Control, Communications, Computers, Intelligence, Surveillance and Reconnaissance) systems for accurate synchronisation of various data and actions across radar, electronic warfare, command centres, weapon systems, etc.

A short note is dedicated to blind quantum computing \cite{Broadbent2009,Fitzsimons2017}. This class of quantum protocols allows for a quantum program to run on a remote quantum computer or quantum computing cloud and retrieve results without the owner knowing what the algorithm or result was. This is valuable when secret computation is needed (e.g. military operation planning or new weapon technology design) and no own quantum computer capability is available.

Distributed quantum computing via the quantum network---see Section~\ref{sec:qunet}---will be important for the military and governmental actors owning quantum computers, to build high-performance quantum computing services or quantum cloud.

A quantum network capable of distributing entanglement can integrate and entangle quantum sensors \cite{Proctor2018} for the purpose of improving the sensitivity of the sensors, reducing errors, and most importantly to perform a global measurement. That provides an advantage in cases where the parameters of interest are global properties of the entire network; for example, when a signal's angle of arrival needs measurement from three sensors, where each measures a signal with a certain amplitude and phase. Afterwards, each sensor's output can be used to estimate the angle of arrival of the signal. Quantum entangle sensors can evaluate this globally. This process can then be improved by machine learning \cite{Xia2021}. 

Quantum protocols for distributed computing agreement \cite{BenOr2005} can have advantageous military application for a swarm of drones, or in general for a herd of autonomous vehicles (AVs). Here, quantum protocols can help achieve agreement between all AVs at the same time scale, independent of their quantity. Nevertheless, open space quantum communication between all rapidly moving AVs will be a challenge that has to be solved first. Note that the first experiment of quantum entanglement distribution from a drone was successfully carried out, recently \cite{Liu2021}.

%...................................................................................
\subsection{Quantum PNT}\label{sec:PNT}

Key points:
\begin{itemize}[noitemsep]
\item All quantum PNT technologies have in common the demand for a highly accurate quantum clock.
\item Quantum inertial navigation could bring few orders of magnitudes higher precision than its classical counterpart.
\item Quantum inertial navigation can be extended by the quantum augmented navigation using quantum magnetic or gravity mapping. 
\item Promising quantum navigation based on Earth's magnetic anomalies.
\end{itemize}

Quantum technologies are expected to significantly improve positioning, navigation and timing (PNT) systems, especially inertial navigation.
Time standards and frequency transfer (TFT) is a fundamental service that provides precise timing for communication, metrology, but also global navigation satellite system (GNSS). Although present TFT systems are well established, the performance of optical atomic or quantum clocks in combination with TFT utilizing quantum networks \cite{Giovannetti2001,Jozsa2000} will keep pace with the increasing demands of the present applications (communication, GNSS, financial sector, radars, electronic warfare systems) and enables new applications (quantum sensing and imaging).

New quantum-based technologies and approaches support the development of sensitive precision instruments for PNT. The quantum advantage will be manifested for GPS denied or challenging operational environments, enabling precise operations. Examples of such environments are underwater and underground, or environments under GPS jamming.

Current GNSS (GPS, GLONASS, Galileo, BeiDou, ...) rely on precise timing provided through multiple atomic clocks in individual satellites that are corrected by the more stable atomic clocks on the ground. The higher precision of the quantum clock will increase the accuracy of positioning and navigation as well. Over the long term, the GNSS satellites should be connected to the quantum internet for timing distribution and clock synchronisation. Chip-size precise mobile clocks could help discover GNSS deception and spoofing \cite{Krawinkel_Schön_2015}.

Some quantum GNSS (not only quantum clock) have been considered and investigated; for instance, interferometric quantum positioned system (QPS) \cite{Giovannetti2001,Bahder:2004pi,Yang2010}.
One of the schemes of QPS \cite{Bahder:2004pi,Yang2010} has a structure similar to the traditional GNSS where there are three baselines, each consisting of two low-orbiting satellites, with the baselines are perpendicular to each other. However, although theoretically the accuracy of positioning is astonishing, significant engineering must be done to design a realistic QPS.

Most of the current navigation relies on GPS, or in general GNSS, which is the most precise available technology for navigation. GNSS technology is prone to jamming, deception, spoofing or GPS-deprived environments such as densely populated areas with high electromagnetic spectrum use. Moreover, for underground or underwater environments, GNSS technology is not available at all. The solution is inertial navigation. 
The problem with classical inertial navigation is its drifting, a loss of precision over time. 
For example, the marine-grade inertial navigation (for ships, submarines and spacecraft) has a drift $1.8$~km/day and navigation grade (for military aircraft) has a drift 1.5~km/hour \cite{EUJRC2020}. In 2014, DARPA started a MTO-PTN project with a goal to reach drift 20~m and 1~ms/hour \cite{DARPA-MTO-PTN}. Even so, some expectations are very high, that quantum inertial navigation will offer error of only approximately hundreds of meters per month \cite{UKLandscape2016,EUJRC2016}.

The full quantum inertial navigation system consists of a quantum gyroscope, accelerometer and atomic/quantum clocks. 
Although the individual sensors required for quantum inertial navigation are tested out of laboratories, it is still challenging to create a complete quantum inertial measurement unit. For navigation for highly mobile platforms, sensors need fast measurement rates of several 100~Hz, or to improve the measurement bandwidth of quantum sensors \cite{EUJRC2020,Bongs2019}. The key component that needs the most improvement is the low-drift rotation sensor.
The classical inertial sensors are based on various principles \cite{ElSheimy2020}. One common chip-size technology is the  MEMS (Micro Electro Mechanical Systems) technology, where MEMS gyroscopes have demonstrated instabilities at level $\sim 10^{-7}$~rad$\cdot$s$^{-1}$ that is suitable for military applications \cite{Geiger2020}. The instability limit for the best current cold-atom gyroscopes is about $\sim 10^{-9}-10^{-10}$~rad$\cdot$s$^{-1}$ (at integration time 1000~s) \cite{Savoie2018}. The uncertainty is in the precision of the in field-deployable quantum sensors in comparison to the presented laboratory experiments' precision.
The intermediate step between classical and quantum inertial navigation can be a hybrid system fusing the outputs of classical and quantum accelerometers \cite{wang2021enhancing}.
With the size of the quantum inertial navigation device decreasing to chip size, its deployment can be expected on smaller vehicles, especially unmanned autonomous vehicles or missiles. However, the miniaturisation we can reach is unknown. There are many doubts about chip-sized quantum inertial navigation. It is certainly a next-generation technology, although a very big challenge.

Currently, the individual elements, such as gyroscope or accelerometer, are also tested on various platforms; for instance, on board an aircraft \cite{Battelier2016}, or more recently a \cite{QISplane}.

For many years, the US National Oceanic and Atmospheric Administration (NOAA) were mapping the Earth's magnetic anomaly and creating a magnetic anomalies map. Using sensitive quantum magnetometers in combination with Earth's magnetic anomaly map is another way to realise quantum non-GNSS navigation \cite{Canciani2017,MagInerNav}. 

Gravitational map matching \cite{Moryl1998} works on a similar principle, and one can expect improved performance using the quantum gravimeter.
Together, quantum gravimeter and magnetometer could be a basis for a submarine quantum augmented navigation, especially in undersea canyons, wrinkled seabeds, or littoral environments.

In general, quantum inertial navigation or augmented navigation has vast potential, since there is no need for GPS, infra or radar navigation and it is not susceptible to jamming, or in general to electronic warfare attacks. However, the claim of `no need for GPS' is not quite accurate. These systems will always need some external input on their initial position, most probably from GNSS.

%...................................................................................
\subsection{Quantum ISTAR}

Key points:
\begin{itemize}[noitemsep]
\item Intense involvement of quantum computing to gather and process information.
\item Desired deployment on low-orbit satellites, but the resolution is questionable.
\item Vast applications for undersea operations.
\item Expected advanced underground surveillance with uncertain resolution.
\item New type of 3D, low-light or low-SNR quantum vision devices.
\end{itemize}

ISTAR (intelligence, surveillance, target acquisition and reconnaissance) is a crucial capability of a modern army for precise operations.
Quantum technologies have the potential to dramatically improve situational awareness of multi-domain battlefields.

In general, a large impact can be expected from quantum computing that will help with acquiring new intelligence data, processing Big Data from surveillance and reconnaissance and identifying targets using quantum ML/AI \cite{Wilson2020,Spiegeleire2017}. 

Apart from the processing part of ISTAR, dramatic advancement can be expected from quantum sensing placed on individual land/sea/aerial vehicles and low-orbit satellites.

Quantum gravimeters and gravitational gradiometers promise high accuracy that can improve or introduce new applications:
geophysics study, seismology, archaeology, minerals (fissile material or precious metals) and oil detection, underground scanning and precise georeferencing and topographical mappings (e.g. of the seabed for underwater navigation) \cite{IDA2019}.

Another significant type of sensing is quantum magnetometry. The applications of quantum magnetometry are partially overlapped by applications for quantum gravimetry, thus introducing new applications:
Earth's magnetic field including magnetic anomalies, local magnetic anomalies due to the presence, such as metallic objects (submarines, mines, etc.), or weak biological magnetic signals (applications mainly for medical purposes) \cite{IDA2019}.

The third field interesting for ISTAR is quantum imaging. Quantum imaging offers plenty of diverse applications; for example, quantum radar (see Section~\ref{sec:quantumradar}), imaging devices for medicine, 3D camera, stealth rangefinder, etc.

The potential quantum computing applications in ISR and situational awareness are described in Sec.~\ref{sec:QCcap}.

%...................................................................................
\subsubsection{Quantum Earth's Surface and Underground Surveillance}\label{sec:ISTARmag}

Quantum sensing based on magnetometry, gravimetry and gravity gradiometry at the first level helps with the study of continents and sea surface, including underground changes of natural origin. Both magnetic anomaly and gravity-based sensing provide a different picture of the Earth's surface.
The Earth is very inhomogeneous (ocean, rocks, caves, metallic minerals, ...), including the massive constructions or vehicles made by people which generate a unique gravitational (depending on the mass) and magnetic (depending on metallic composition) footprint. 

The discussed quantum sensing technologies---magnetometry, gravimetry and gravity gradiometry---can reach very high precision, at least in the laboratory. For example, the precision of absolute gravimetry out of the laboratory is about $1~\mu$Gal (10 nm$\cdot$s$^{-2}$) \cite{Mnoret2018}. Note that the sensitivity of $3.1~\mu$Gal corresponds to a sensitivity per centimetre of height above the Earth's surface.
However, the problem is the spatial resolution that usually is anti-correlated with the sensitivity (higher sensitivity is at the cost of lower spatial resolution and vice versa). Spatial resolution and sensitivity are the critical attributes that define what you will recognise (large-scale natural changes or small underground structures) and from what distance (from the ground, drone or satellite-based measurement).
Examples of the current spatial resolution are about 100~km \cite{Bidel2018} for satellite-borne gravity gradiometer or 16~km \cite{Sandwell2014} additional width using radar satellite altimetry (for sea areas), or 5~km \cite{Bidel2020} for airborne gravimetry. For more information, see e.g. \cite{UKLandscape2016}.

For many quantum sensing applications, it would be essential to place sensors on low Earth orbit (LEO) satellites \cite{Travagnin2020}. However, the current sensitivity and spatial resolution allow only the applications for Earth monitoring (mapping resources such as water or oil, earthquake or tsunami detection).

Apart from low-orbit satellites, the mentioned quantum sensors are considered for deployment on airborne, sea or ground vehicle platforms. 
Nowadays, quantum sensing experiments are performed outside the laboratory environment, such as in a truck \cite{Mahadeswaraswamy2009}, on drones and aeroplanes \cite{Geiger2011,Perkins2016} or aboard ships \cite{Bidel2018}.
For example, the quantum gravimeter could be mounted on drones to search for human-made structures such as tunnels used to smuggle drugs \cite{Perkins2016}. Placing quantum sensing devices on a drone (this may be an unmanned aerial vehicle (UAV), Unmanned Surface Vessel (USV), Remotely Operated Vehicle (ROV) or unmanned underwater vessel (UUV)) needs more engineering to reach the best sensitivity, resolution and operability simultaneously.

Low-resolution quantum sensing could be used for precise georeferencing and topographical mappings to help with underwater navigation or mission planning in rugged terrain. Also, the detection of new minerals and oil fields can become a new centre of interest, especially under the seabed \cite{guo_cross-border_2005}. This can be a source of international friction, despite the fact that borders are clear in most cases.

High-resolution quantum magnetic and gravity sensing \cite{Battersby2016,GravityPioneer,Marmugi2017,Bidel2018} is considered in numerous reports and articles \cite{IDA2019,yasmin_application_2018,Bond2019,Kumar2004,streland_going_2003,Battersby2016} to be able to:
detect camouflaged vehicles or aircraft;
effectively search for a fleet of ships or individual ships from LEO;
detect underground structures such as caves, tunnels, underground bunkers, research facilities and missile silos;
localise buried unexploded objects (landmines, underwater mines and improvised explosive devices);
achieve through-wall detection of rotating machinery. 

However, note again that it is highly uncertain where the technical limits are and whether the mentioned quantum gravimetry and magnetometry applications will reach such sensitivity and resolution (especially for using from LEO) as to realise all the aforementioned ideas.
Quantum sensors will be delivered to the market in many generations, each with better sensitivity and resolution and lower SWaP, allowing more extensive deployment and application.

%...................................................................................
\subsubsection{Quantum Imaging Systems}

Besides quantum radar and lidar (see Section~\ref{sec:quantumradar}), there are other military-related applications of quantum imaging. In general, all-weather, day-night tactical sensing for ISTAR for long/short-range, active/passive regime, invisible/stealth using EO/IR/THz/RF frequencies features and advantages are considered. Quantum imaging systems can use various techniques and quantum protocols; for example, SPAD, quantum ghost imaging, sub-shot-noise imaging, or quantum illumination as was described in Section~\ref{sec:qimag}.
In general, it is not a problem to construct quantum imaging systems of small sizes. The critical parameters are the flux of the single-photon/entangled photon emitter or the single-photon detection resolution and sensitivity. Moreover, a large-scale deployment of a quantum imaging system with high photon flux will require powerful processing that can limit the system deployability and performance.

Quantum 3D cameras exploiting quantum entanglement and photon-number correlations will introduce fast 3D imaging with unprecedented depth of focus with low noise aiming at sub-shot noise or long-range performance. This capability can be used to inspect and detect deviation or structural cracks on jets, satellites and other sensitive military technology. Long-range 3D imaging from UAV can be used for reconnaissance and to explore mission destination or hostile facilities and equipment.

Another commercially available technology is quantum gas sensors \cite{QMLTech}. Technically it is a single-photon quantum lidar calibrated to detect methane leakage. The next prepared product is a multiple gas detector able to also detect carbon dioxide (CO$_2$). With proper improvement and calibration, it could serve for human presence detection, too. 

A specific feature at short range is the possibility of behind-the-corner or out of the line-of-sight visibility, \cite{Gariepy2015}. These methods can help to locate and recover trapped people, people in hostage situations or to improve automated driving by detecting incoming vehicles from around a corner.

Quantum imaging can serve as a low-light or low-SNR vision device; for example, in an environment such as cloudy water, fog, dust, smoke, jungle foliage or in the nighttime, leading to an advantage. Low-SNR quantum imaging could help in target detection, classification and identification with low signal-to-noise ratios or concealed visible signatures and potentially counter adversaries' camouflage or other target-deception techniques.
Quantum imaging will be very useful for helicopter pilots when landing in dusty, foggy or smoky environments \cite{NQIT2018}.

One significant product will be a quantum rangefinder \cite{Cohen2019,Frick2020}. Conventional rangefinders use a bright laser and can be easily detected by the target. A quantum rangefinder will be indistinguishable from the background both temporally and spectrally when viewed from the target. In other words, the quantum rangefinder will be invisible and stealthy, including at night time, whereas the classical rangefinder can be visible to the target or others.

Under some circumstances, quantum ghost imaging can play the role of quantum lidar \cite{Hardy2013}, especially when the target does not move or moves very slowly and infinite depth of focus is required for 3D imaging.

%...................................................................................
\subsection{Quantum Electronic Warfare}\label{sec:quantumEW}

Key points:
\begin{itemize}[noitemsep]
\item Enhancement of current EW by smaller universal quantum antennas, precise timing and advanced RF spectrum analysers.
\item The problem with detection of quantum channels.
\item When the quantum channel is localised, several types of attacks are considered and developed.
\end{itemize}

Quantum electronic warfare (EW) can be divided into quantum-enhanced classical EW and quantum EW focusing on countermeasures, counter-countermeasure and support against quantum channels. By a quantum channel is meant any transfer of photons carrying quantum information for quantum internet, quantum radar or another quantum system that uses the free-space or optical fibres channel. 

Classical EW systems for electronic support measures can benefit from the quantum antenna. Quantum antenna based on Rydberg atoms can offer a small size independent of the measured signal wavelength (frequency) \cite{Facon2016,Cox2018}. This means that even for low-frequency (MHz to kHz \cite{Meyer2020,Meyer2021}) signal interception a few-micrometres of quantum antenna is sufficient. There can be an array of quantum antennas for multi-frequency measurement for different bandwidths or one antenna dynamically changing  bandwidth according to the interest. 
Moreover, Rydberg atoms-based antennas can measure both AM and FM signals, offer self-calibration, and measure both weak and very strong fields and detect the angle-of-arrival \cite{robinson2021determining}.
In the future, quantum antennas could look like an array (matrix) of Rydberg atom cells. Different cells can measure different signals, and in the joint measurement of two or more cells, the angle-of-arrival of the signal could be determined.
The weakest aspect of such antennas is the cryogenics required for cooling Rydberg atoms that need to be scaled down to an acceptable size. 
In general, quantum RF sensors are a key enabler for advanced (LPD/LPI\footnote{Low Probability of Intercept/Low Probability of Detection (LPI/LPD)}) communications, over-horizon directional RF, resistance to RF interference and jamming, RF direction finding, or RF-THz imaging.
As an example, an arrayed quantum RF sensor is developed as a potential upgrade for fighter F-35 \cite{ColdQuanta-RF}.

Classical EW can also benefit from quantum computing, offering improved RF spectrum analysers for electronic warfare where quantum optimisations and quantum ML/AI techniques can be applied. Higher effectiveness can be reached by the processing and analysing directly of quantum data \cite{Dunjko2016} from RF quantum sensors (Rydberg atoms, NV centres), where the impact of a quantum computer can be more significant.
Moreover, other quantum-based solutions and approaches are under development, such as NV centre based RF spectrum analysis or SHB based rainbow analyser
\cite{Baili2009}.

The current EW systems will also benefit from quantum timing. Quantum timing can enhance capabilities such as signals intelligence, counter-DRFM (digital radio frequency memory) and other EW systems that require precise timing; for instance, counter-radar jamming capabilities.

The other area of quantum EW will be signals intelligence (SIGINT) and communications intelligence (COMINT) (detecting, intercepting, identifying, locating) and quantum electronic attack (jamming, deception, use of direct energy weapons). 
Quantum channels (for quantum communication or quantum imaging) have specific characteristics.
First, the simple signal interception is problematic because the quantum data are carried by individual quanta, and their interception can be easily detected. Second, typical quantum imaging technologies use a low signal-to-noise ratio, which means that it is challenging to recognise signal and noise without extra knowledge. 
Third, coherent photons, usually used as a signal, behave like a laser that is very focused. Finding such a quantum signal without knowing the position of at least one party is very challenging.    
These characteristics make the classical EW obsolete and blind against quantum channels.

The situation is difficult even for potential quantum electronic warfare systems, since it is open to question whether it will be possible to detect the presence of a quantum (free-space) channel. This will require the development of quantum analogy of laser warning receivers \cite{Pietrzak2003}. For quantum EW, it will be critical to get intel on the position of one or both parties using the quantum channel.

Classical EW would intercept and eavesdrop on the free-space classical channel. However, this is not possible for the quantum channel where it would be detected promptly. One possible attack is a man-in-the-middle type attack \cite{Fei2018,Vergoossen2019}, since the early quantum network parties can have a problem with authentication or trusted repeaters. 
Other types of attacks are considered at the quantum physics level; for example, a photon number splitting attack relies on utilising coherent laser pulses for the quantum channel \cite{Brassard2000} or the Trojan-horse attacks \cite{Jain2014}, or the collecting of scattered light and its detection \cite{Lee2019}.
However, these types of attacks are very sophisticated, and their practicability, for example in space, is uncertain.

It is more probable that the quantum EW attack will be just a type of denial of service, where the quantum channel is intercepted, leading to stoppage of use of the channel. Another possibility is the sophisticated jamming of the receivers on one or both sides, leading to enormous noise. 
When the position of the receiver or transmitter is known, another countermeasure of the classical EW is to make use of directed energy weapons such as laser, leading to damage or destruction of sensors. Such an attack also could help eavesdroppers \cite{Bugge2014}.

In general, new approaches and methods will need to be developed to realise the capabilities of quantum electronic warfare and address the corresponding requirements.

%...................................................................................
\subsection{Quantum Radar and Lidar}\label{sec:quantumradar}

Key points:
\begin{itemize}[noitemsep]
\item Long-range surveillance quantum radar is unlikely with existing quantum microwave technology.
\item Possible applications in the optical regime - quantum lidar.
\item Quantum radar could be used for space warfare.
\end{itemize}

The perception of the quantum radar topic \cite{lanzagorta2011quantum,torromé2021introduction,Shapiro2020} is affected by the hype in the media claiming quantum radar development in China \cite{CETC2016,AviWeek2018} or by optimistic laboratory experiments. Indeed, the theoretical advantages and features of quantum radar are significant (some of them depend on individual quantum protocols):
\begin{itemize}[noitemsep,topsep=-1em,after=\vspace{\baselineskip}]
\item[-] Higher resistance to noise---that is, better SNR (signal-to-noise ratio)---higher resistance to jamming and other electronic warfare countermeasures;
\item[-] Based on individual photons; that is, the output signal power is so low that it will be invisible to electronic warfare measures;
\item[-] Target illumination; that is, a radar allowing identification of the target.
\end{itemize}
Based on the list of unique quantum radar features, it could be a powerfully disruptive technology that could change the rules of modern warfare.
Therefore, attention is being paid to this topic internationally, despite the immaturity of the technology, and the many doubts about whether the quantum radar could work as the standard primary surveillance radar.

Moreover, many people immediately imagine quantum radar as a long-range surveillance radar with a range of hundreds of kilometres, whereas such an application of quantum radar seems unlikely \cite{Daum2020,Karsa2020}.
Such an optimal, long-term surveillance quantum radar would be extremely expensive (many orders of magnitude higher than the classical radar cost for any range) \cite{Daum2020}, and it would still not fulfil all the advantages and features listed above.

Briefly, the practical problems are the following \cite{Daum2020}. Quantum radar too is subject to the radar equation, where the received power is lost with the distance's fourth power.
In parallel, to keep the quantum advantage, it is desirable to have one or fewer photons per mode. In summary, the relatively high power made of low-photon modes in the microwave regime is needed to be generated. This requires a lot of quantum signal generators, cryogenics, large antenna sizes, etc. All this leads to extremely high cost, and impractical design \cite{Daum2020,Luong2020}. Scientists need to come up with more practical quantum microwave technology to overcome these difficulties. 

Apart from the high price, scepticism also remains about the detection of stealthy targets or jamming resistance. Quantum radar can be advantageous against a barrage jammer, but not necessarily against a DRFM or other smart jammer \cite{Daum2020}. 
In summary, the long-range surveillance quantum radar is unlikely to be achieved even as a long-term prospect. For its realisation, one would need to evolve new technology allowing smaller cryogenics, RF quantum emitter working at a higher temperature or more efficient cryogenics cooling, and a more powerful emitter (high rate of low photon pulses). Note that even if the room-temperature superconducting materials were developed, it would not help in the Josephson parametric amplifier (JPA) method of entangled microwave photon generation \cite{Luong2020b}. Nevertheless, JPA is not the only method to obtain entangled microwave photons \cite{Luong2020}.
It is not entirely impossible that a new theory and designs of quantum radar will be discovered in the future.
The long-range surveillance quantum radar described above would suffer from large size, weight and power consumption, and it is questionable if such a radar would be stealthy \cite{Daum2020}.

Another problem is the ranging in the case of quantum illumination (QI) protocol. QI protocol requires knowledge of the target in advance, and therefore it requires some extension for ranging, whether classical or quantum \cite{ASPI2017}. 

For several years, it was believed that the quantum radar cross section (RCS) is higher than the RCS of classical radars \cite{Lanzagorta2010,Brandsema2016}.
A new precise study of quantum RCS \cite{Brandsema2020} shows that the previously claimed advantage of quantum RCS over the classical RCS results from erroneous approximation. Quantum and classical RCS seem to be equal, at the moment.

Another approach can be the quantum-enhanced noise radar \cite{Chang2019,Lukin2020,Luong2020}.
Noise radar uses noise waveform as a transmission signal, and detection is based on the correlation between the transmitted signal and the received noise waveform radar returns. The advantage is the low probability of interception (LPI), being nearly undetectable by today's intercept receivers.
The quantum noise radar design needs more study to see practical applicability. However, a potential use here is especially for the microwave regime.

Still, the current theory and research have applications in the radar sector, especially that which uses the optical or near-optical photons; that is, quantum lidar. Here, a short-range quantum lidar could be used for target illumination at short distances. Experiments with single-photon imaging were demonstrated from 10 \cite{Pawlikowska2017} to 45 km \cite{Li2020}. In this range, quantum lidar could operate as an anti-drone surveillance radar or as part of a SHORAD (Short Range Air Defense) complex.

Space can be another example of an advantageous environment for quantum radar/lidar \cite{Lanzagorta2015} which is low noise for the optical regime, and it even almost eliminates the decoherence problem in the case of entangled photons.
For example, Raytheon performs simulations of the quantum radar in the optical regime for space domain \cite{koblick_space-based_2020,Raytheon2020}. The idea is to place a quantum radar on a satellite and detect small satellites that are difficult to detect because of their small cross-sectional area, reflectivity, and environmental lighting conditions. 
The deployment of quantum radar/lidar for the space environment can provide almost all the advantages listed above.

A small note is dedicated here to quantum-enhanced radar. Classical radar can be equipped with an atomic or quantum clock. Such quantum-enhanced radars show high precision and reduced noise, and thus demonstrate an advantage in detecting small, slow-moving objects such as drones \cite{QuantEnhancRadar}.

%...................................................................................
\subsection{Quantum Underwater Warfare}

Key points:
\begin{itemize}[noitemsep]
\item Submarines can be one of the first adopters of quantum inertial navigation.
\item Quantum magnetometers as the main tool for detection of submarines or underwater mines.
\end{itemize}

Quantum technologies can significantly interfere in underwater warfare, with enhanced magnetic detection of a submarine or underwater mines, novel inertial submarine navigation and quantum-enhanced precise sonars.  In general, in the maritime environment, sensing based on quantum photo-detectors, radar, lidar, magnetometers, or gravimeters can be applied \cite{Lanzagorta2015}. 
For a general overview of the implications of quantum technology for nuclear weapon submarines' near invulnerability, see \cite{KubiakSubmarines}. 

Submarines and other underwater vehicles will benefit from quantum inertial navigation described in Section~\ref{sec:PNT} about PNT. Large submarines can probably be one of the first adopters of quantum inertial navigation because they can afford to install larger quantum devices, including cryogenics cooling. Moreover, sensitive quantum magnetometers and gravimeters can help map surroundings such as an undersea canyon, icebergs and a wrinkled sea bottom without using sonar that can be easily detected.
An example of another type of inertial navigation especially suitable for underwater arctic navigation is based on quantum imaging \cite{Lanzagorta2017}.

The basic tool for anti-submarine warfare could be the quantum magnetometer. 
Researchers anticipate that the SQUID magnetometers in particular could detect a submarine from 6 kilometres away, with still improving noise suppression \cite{Submarine6km,NIsubmarineWarfare}. Note that the current classical magnetic anomaly detectors, usually mounted on a helicopter or a plane, have a range of only hundreds of meters. An array of quantum magnetometers, such as along the coast, could cover significant areas, leading to denial area for submarines. Moreover, an array of quantum magnetometers seems to work better with more suppressed noise. 

Quantum magnetometers can also be used to detect underwater mines using, for instance, an unmanned underwater vessel \cite{Kumar2004}.

However, the main discussion is about the detection range, sensitivity, etc., as in Sec.~\ref{sec:ISTARmag}. Even other underwater domain technology such as sonar offers longer detection range \cite{Bond2019}.
It was also pointed out in \cite{KubiakSubmarines} that quantum technologies will have little impact on SSBN (ballistic missile submarines).
It is possible that quantum magnetometers could work with other sensors to aid in detection, identification and classification of targets \cite{Bond2019}.

%...................................................................................
\subsection{Quantum Space Warfare}

Key points:
\begin{itemize}[noitemsep]
\item Important for long-distance quantum communication.
\item Low Earth orbit will be important for the future deployment of quantum sensing and imaging technologies.
\item Space warfare will lead to new quantum radar/lidar and quantum electronic warfare technologies for deployment in space.
\end{itemize}

The space domain is gaining in importance and will be an important battlefield used by advanced countries. Space used to be a place mainly for satellites for navigation, mapping, communication and surveillance, often for military purposes. Nowadays, space is becoming more weaponised \cite{SpaceWeaponization}; for example, satellites with laser weapons or `kamikaze' satellites are placed in Earth orbit, and anti-satellite warfare is growing in parallel.
Another surging problem is the amount of space garbage, with the number of satellites estimated at 2,200 and several more planned to be released \cite{Sattelites2019}.

Space also will be key for placing quantum sensing and communication technology in satellites \cite{QTSpace,Jia2014,Oi2017,sidhu2021advances,pirandola2021satellite}, as well as for space countermeasures.

For many quantum technology applications described in previous sections, it would be desirable to place quantum sensing technology such as quantum gravimeter, gravity gradiometer or magnetometer on satellites in Earth orbit, especially the low one (LEO). Such applications are in development; for example, a low-power quantum gravity sensing device that can be deployed in space on board a small satellite for accurately mapping resources or to aid in assessing the impact of natural disasters \cite{NASAchallenge}. However, such an application does not require too high spatial resolution. See Sec.~\ref{sec:ISTARmag} for a detailed discussion.
The same applies to satellite-based quantum imaging. For example, China claimed the development of a spy satellite that uses ghost imaging technology \cite{SCMPghost}. However, what spatial resolution it has is uncertain. Nevertheless, quantum ghost imaging would have the advantage of being usable in cloudy, foggy weather or at night as well.

On the other hand, utilisation of satellites for quantum communication has already been demonstrated \cite{Yin2017,Liao2017}. Satellite-based quantum communication will be essential for the near-term integrated quantum network at long distances \cite{Chen2021}. 
The present quantum communication satellites suffer from the same problems as trusted repeaters for optic fibre channel. In fact, present quantum satellites are trusted repeaters.  
The issue with trusted repeaters is that they keep the doors open to possible cyber attacks on the satellite control system. A better security situation is with the presently demonstrated MDI-QKD protocol \cite{Cao2020}, where the central point works as a repeater or switch, but in a safe regime, and later with quantum repeaters.
For a space quantum communication overview, see \cite{sidhu2021advances,pirandola2021satellite}.

A new required military capability will be technology to detect other satellites, space-borne objects, space garbage and track them. classical radars are used for this purpose; for instance, the Space Fence project as part of the US Space Surveillance Network \cite{SpaceFence}. However, most of these space surveillance radars have problems with objects with a size of about 10~cm and smaller \cite{Sattelites2019} (in the case of Space Fence, the minimal size is about 5 cm), and another problem is the capacity, as to how many objects they can track.
This is the case with most of the space garbage that is only a few centimetres in size.
Instead of classical radar, quantum radar or lidar is considered \cite{Lanzagorta2015,ASPI2017,Raytheon2020} as an alternative. 
Specifically for the space environment, the quantum radar in optical regime is considered \cite{Raytheon2020}, since the optical photons do not suffer from losses such as in the atmosphere.
Space quantum radar can offer most of the advantages of quantum radar as described in Section~\ref{sec:quantumradar}, including stealth.
According to simulations \cite{Raytheon2020}, quantum radar in space can offer at least one order of magnitude higher detection sensitivity
and object tracking sensitivity in space in comparison with GEODSS (Ground-based Electro-Optical Deep Space Surveillance).
Space quantum radar would be very useful for tracking small, dark and fast objects, such as satellites, space garbage or meteoroids.

The increasing presence of quantum sensing and communication devices in space will lead to increased interest in quantum electronic warfare as described in Section~\ref{sec:quantumEW}.

%...................................................................................
\subsection{Chemical and Biological Simulations and Detection}

Key points:
\begin{itemize}[noitemsep]
\item $\sim200$ qubits are sufficient to carry out chemical quantum simulation research.
\item The capability of achieving more complex simulations increases with the number of logical qubits.
\item Chemical detection in the air or in samples.
\item Suitable for detecting explosives and chemical warfare agents.
\end{itemize}

The defence-related chemical and biological simulations are primarily interesting for the military and national laboratories, the chemical defence industry or CBRN (Chemical, Biological, Radiological and Nuclear) defence forces. Research on new drugs and chemical substances based on quantum simulations will require an advanced quantum computer, classical computing facility and quantum-chemical experts.
The quantum simulations for chemical and biological chemical warfare agents, in principle, have the same requirements as civil research, such as the already ongoing protein folding, nitrogen fixation and peptides research.

The number of required qubits depends on the number of spatial basis functions (various basis sets exist, e.g., STO-3G, 6-31G or cc-pVTZ); for example, using the 6-31G basis, the Benzene and Caffeine molecules can be simulated by approx. 140 and 340 qubits, respectively \cite{AspuruGuzik2005}.
Then, the Sarin molecule simulation, for instance, requires about 250 qubits. Based on quantum computer roadmaps \cite{Google1million,China:1million} and logical qubit requirements, one can come to 100 logical qubits in 10 years, but probably earlier with more effective error corrections and error-resisting qubits. This is sufficient for medium-sized molecule simulations.

The threat could be the design and precise simulation of structures and the chemical properties of new small- to medium-sized molecules that could play the role of chemical warfare agents similar to, for example, Cyanogen, Phosgene, Cyanogen chloride, Sarin or Yperit.
On the other hand, in general the same knowledge can also be used for CBRN countermeasures and new detection technique development.

The research on protein folding, DNA and RNA exploration, such as motifs identification, Genome-wide association studies and De novo structure prediction \cite{emani2019quantum} could impact the research on biological agents as well \cite{Vignard:2000}. However, more detailed studies are needed to assess the real threat from quantum simulations.

Photoacoustic detection with quantum cascade laser will be effective as a chemical detector. For example, quantum chemical detectors can detect TNT and triacetone triperoxide elements used in improvised explosive devices (IED) that are a common weapon used in asymmetric conflicts. The same system for detecting Acetone can be used to discover baggage and passengers with explosives boarding aircraft.  
In general, quantum chemical detection can be used against chemical warfare agents or toxic industrial chemicals \cite{Kumar2007,Li2015}.

In the mid- to long term, such detectors can be placed on autonomous drones or ground vehicles that are inspecting an area \cite{QCDdrone}.

%...................................................................................
\subsection{New Material Design}

Key points:
\begin{itemize}[noitemsep]
\item General research impacts; for example, room-temperature superconducting allowing the highly precise SQUID magnetometers to operate without cooling can have a remarkable impact on military quantum technology applications.
\item Defence industry research on camouflage, stealth, ultra-hard armour or high-temperature tolerance material.
\end{itemize}

Modern science is developing new materials, metamaterials, sometimes called quantum material, by exploiting the quantum mechanical properties (e.g. graphene, topological insulator). Material as a quantum system can be simulated by a quantum computer; for example, the electronic structure of the material. 
The considered applications can be, for instance, the room-temperature superconductor, better batteries and improvement of specific material features. 

To explain in greater detail, the room-temperature superconductivity material, for example, exploits superconductivity at high temperatures \cite{Cirac2012}. That would allow building Josephson junctions, usually used as the building blocks of SQUIDs or superconducting qubits.
So far, cooling near absolute zero is required. It is expected that a quantum computer with about 70 logical qubits \cite{AWS2020} could be sufficient for the basic research on high-temperature superconductors.

For the defence industry, opportunities for research on new materials such as better camouflage, stealth (electromagnetic absorption), ultra-hard armour or high-temperature tolerance material design are considered without any details being revealed.\footnote{Usually mentioned in public news articles or interviews but without details or references.}

%...................................................................................
\subsection{Brain imaging and human-machine interfacing}
Key points:
\begin{itemize}[noitemsep]
\item Quantum enabled magneto-encephalography
\item Enhanced human-machine interfacing
\end{itemize}

MEG (magneto-encephalography) scanner is a medical imaging system that visualises what the brain is doing by measuring the magnetic fields generated by current flowing through neuronal assemblies.
Quantum magnetometers---based, for instance, on optically pumped magnetometers \cite{Tierney2019}---can enable high-resolution magnetoencephalography for real-time brain activity imaging.
This technology is safe and non-invasive, and is already laboratory tested. The technology itself is small, and wearable \cite{Tierney2019}.

In the near term, quantum MEG could be a part of a soldier's helmet for continuous and remote medical monitoring and diagnosis in case of injury.
The long-term expectations include enhanced human--machine interfacing, i.e. practical non-invasive cognitive communication with machines and autonomous systems \cite{AusQRoadmap}.

%---------------------------------------------------------------------------
%---------------------------------------------------------------------------
%---------------------------------------------------------------------------
\section{Optimism versus Pessimism}\label{sec:optipess}

Many of the quantum technology military applications mentioned above sound very optimistic and can drive exaggerated expectations. Some applications are taken from various reports and newspapers or magazine articles, wherein the author could have overestimated the quantum technology transfer from the laboratory to the battlefield or been influenced by general quantum technology hype \cite{Biercuk2017hype}.
It is especially important to avoid exaggerated expectations when the topic concerns national security or defence. 
This issue has been described in \cite{Biercuk2020war}.

Quantum technology military applications described above are based on public-domain, state-of-the-art research supplemented by various reports and newspaper or magazine articles about defence applications. 
Critical remarks on their feasibility are not given for several technologies, since there is no public information on the same.
In these cases, the reader should be more careful and critical until more detailed studies are available.

On the other hand, it is known that big defence corporations and national defence laboratories have had quantum research and development programmes for several years. However, only some detailed information is publicly communicated. The opposite extreme seems to include announcements, such as from China \cite{CETC2016,AviWeek2018,SCMPghost,Submarine6km}, where it is difficult to disentangle the real research advancement from the state's strategic propaganda \cite{CNAS2018}.

For many of the mentioned quantum technologies, only a laboratory proof of concept has been provided so far.
The decisive factors determining whether the quantum technologies will be applied outside the laboratory to general use are component miniaturisation and susceptibility to interference.
These improvements must not be made at the expense of sensitivity, resolution and functionality.
Another decisive factor in real deployment is the price of the technology.

In conclusion, considering the advancements in quantum technology research and in supportive systems, such as laser and cryogenic cooling miniaturisation in the last few years, it is reasonable to be optimistic rather than pessimistic about the future quantum technology military applications (from the perspective of military or governmental actors). One needs to be careful about the real capabilities in operational deployment, to see whether they fulfil the requirements and if the price-performance ratio justifies acquisition and deployment.

%---------------------------------------------------------------------------
%---------------------------------------------------------------------------
%---------------------------------------------------------------------------
\section{Quantum Warfare Consequences and Challenges}\label{sec:consandchall}

The development, acquisition and deployment of quantum technologies for military application will raise new, related challenges. 
The concept of quantum warfare will impose new demands on military strategy, tactics and doctrines, on ethics and disarmament activities and on technical realisation and deployment.
Studies should be conducted to understand the issues, implications, threats and choices that arise from the development of quantum technologies, and not only for military application.

%---------------------------------------------------------------------------
\subsection{Military Consequences and Challenges}

Quantum technologies in military applications have the potential to sharpen the present capabilities, such as by providing more precise navigation, ultra-secure communication or advanced ISTAR and computing capabilities. In general, quantum warfare will require an update, modification or creation of new military doctrines, military scenarios and plans to develop and acquire new techniques and weapons for the quantum age.

Before this, the development of technology policies and strategies is needed to respond to the strategic ambitions of individual actors \cite{DTF2019}.
National technology policies and strategies should include, for example, the research of national quantum technology resources (universities, laboratories and corporations) and markets, the state of development and feasibility studies and the military and security threat and potential assessments, such as \cite{KubiakSubmarines}. 

The monitoring of quantum technology evolution and adaptation is essential to avoid technological surprises due to neighbouring or potentially hostile countries. Quantum warfare monitoring is essential even if the quantum technology is beyond their financial, research or technological capabilities for some countries. Therefore, all modern armies should be interested in the possible impacts of quantum warfare.

The national trade and export policies are also important. For example, the European Union has declared quantum computing as an emerging technology of global strategic importance and is considering more restricted access to the research programme named Horizon Europe \cite{Horizon2021}. Further, China has prohibited the export of cryptography technology, including quantum cryptography \cite{ChinaBan}.

Another topic is the careful communication of significant quantum advantages along with allies, especially in the quantum ISTAR and quantum cyber capabilities, which can reveal military secrets, such as classified files, nuclear submarines' positions or underground facilities. A significant disruption of the balance of power could upset allies as well as neutral or hostile players \cite{NQIT2018}.

%---------------------------------------------------------------------------
\subsection{Peace and Ethics Consequences and Challenges}

To date, the military applications of quantum technologies mapped in Sec.~\ref{sec:warfare} do not introduce new weapons even as they sharpen the existing military technology; for instance, by developing more precise sensing and navigation, new computing capabilities and stronger information security.  
Nevertheless, the question if quantum technologies, especially for military applications, will be good or bad for world peace is relevant.  

Already various calls for ethical guidelines for quantum computing \cite{Castellanos_2021,Khan2021,QuantumEthicsWEF} have appeared, wherein ethical concerns, such as human DNA manipulation, creation of new materials for war and intrusive AI are mentioned \cite{Khan2021}.

Despite the fact that quantum technologies do not generate new weapons, their improvement of present military technology will sharpen such capabilities, shortening the time for an attack, warning and decision making. 
Consequently, quantum technologies can make the use of force more likely even while reducing individual risk \cite{Pfaff_2020}, and thus make war more probable \cite{Altmann_2020a,Altmann_2020b}.

The preventive arms control of generic dual-use technologies such as quantum technologies will be more difficult because they can be used for civilian applications too, such as in quantum sensing for medicine. An analogy with nanotechnologies has been made \cite{Altmann_2005}.
Export controls to prevent or slow down proliferation and military use by other countries or non-state groups are the most likely way to attempt to reduce any threat posed by quantum technologies \cite{Altmann_2020b}.

Specifically, quantum computing research and development is very expensive. However, the goal is to develop a technology that allows simple and reliable qubit production. This can lead to cheaper, more widely distributed and accessible technology for actors with fewer skills, which is a trait of upcoming problematic military technology \cite{Altmann_2020b}.

%---------------------------------------------------------------------------
\subsection{Technical Consequences and Challenges}

The transfer of successful laboratory proofs of concept to real `outside' application faces many technical and technological challenges, such as miniaturisation and operability that is not at the expense of laboratory-achieved sensitivity and resolution. Further, there are other related technical challenges.

A significant problem could be the quantum workforce. The quantum workforce does not need to comprise physicists or scientists with a doctorate. However, they should be quantum engineers with knowledge of quantum information science and a quantum technology overview who can understand and be able to process and evaluate the outgoing data from quantum sensors, computers and communications. 
Presently, an existing quantum ecosystem is growing continuously, and this ecosystem will require an increasingly larger quantum workforce \cite{VenegasGomez2020}. This requires training and educating new quantum engineers and experts; that is, more universities offering quantum programmes and more students taking them. Besides, it can be even more difficult to get these people to work in the army.
Therefore, the basic principles of quantum information and quantum technology should also appear as part of the curriculum at military colleges of modern armies, where quantum technologies are or will be deployed.

Another technical challenge will be the enormous amount of data. Quantum technologies, through all quantum sensors, quantum imaging, quantum communication and computing, will produce a lot of classical and quantum data that will increase requirements for data transmission, processing and evaluation. 
These requirements should be considered during the planning of C4ISR and quantum infrastructure.

The final challenge will be standardisation. The standardisation process is important for the interoperability of devices manufactured by different producers. Apart from the unification interface and communication protocols, the standardisation process can also include security verification, such as in the post-quantum cryptography standardisation process \cite{Moody2020}. 
Various connected devices in particular (such as nodes, repeaters, switches, fibre channels and open-space channels) can be expected in the case of a quantum network, and it is important to develop and implement some standards that will allow the successful transmission of quantum information.

%---------------------------------------------------------------------------
%---------------------------------------------------------------------------
%---------------------------------------------------------------------------
\section{Conclusion}\label{sec:conclusions}

Quantum technology is an emerging area of technologies that utilise the manipulation and control of individual quanta for multiple applications with the potential to be disruptive. 
Many of these applications are dual-use or are directly used for military purposes. However, individual quantum technologies are at TRLs for military use, from TRL 1 (basic principles observed) to TRL 6 (technology demonstrated in relevant environment).

Quantum technology for military applications will not only offer improvements and new capabilities but will also require the development of new strategies, tactics and policies, assessment of threats to global peace and security and identification of ethics issues. All this is covered by the term `quantum warfare'.

In this report, various quantum technologies at different TRL have been described, focusing on possible utilisation or deployment in the defence sector. 
A precise forecast of quantum technology deployment is not possible, since the transition from the laboratory to real-world applications has not been implemented or is in progress. This raises questions such as whether we will be able to reach a resolution providing a real quantum advantage over the classical systems that are usually significantly cheaper and often in action already. 
Despite the fact that the description of the possible military application of quantum technologies sounds very optimistic, one should be wary of the quantum hype and draw attention to the challenges that lie ahead of the real deployment of quantum technologies for military applications.

Quantum technologies can be expected to have strategic and long-term impacts. Nevertheless, the probability of technological surprises affecting military and defence forces is rather low. The best ways to avoid surprises are cultivating quantum technology knowledge and monitoring quantum technology development and employment. Treating quantum technology with care will play the role of quantum insurance.

\section*{Availability of data and material}
The datasets used and analysed during the current study are available from the corresponding author on reasonable request.

\section*{Authors’ contributions}
As the sole author of the manuscript, MK conceived, designed and performed the analysis and review, and wrote the paper.

\section*{Competing interests}
The author declares that he has no competing interests.

\section*{Acknowledgement}
The author is very grateful for several comments and feedback on the draft, especially by Dr Katarzyna Kubiak, Dr Jürgen Altmann and others.
The author is also grateful for the minor comments to the first preprint and valuable suggestions and journal reviewers' comments.

%---------------------------------------------------------------------------
%---------------------------------------------------------------------------
%---------------------------------------------------------------------------

\printbibliography
\end{document}